\documentclass[12pt,preprint]{aastex}

\newcommand{\funits}{\mbox{\thinspace erg\thinspace s$^{-1}\,$cm$^{-2}\,$\AA$^{-1}\,$}}

\newcommand{\lyb}{\mbox{Ly$\,\beta$}}

\newcommand{\IV}{\mbox{\thinspace{\sc iv}}}
\newcommand{\II}{\mbox{\thinspace{\sc ii}}}
\newcommand{\III}{\mbox{\thinspace{\sc iii}}}
\newcommand{\I}{\mbox{\thinspace{\sc i}}}
\newcommand{\V}{\mbox{\thinspace{\sc v}}}
\newcommand{\VI}{\mbox{\thinspace{\sc vi}}}
\newcommand{\kms}{\mbox{\thinspace km\thinspace s$^{-1}\;$}}
\newcommand{\kmsd}{\mbox{\thinspace km\thinspace s$^{-1}$}}
\newcommand{\gs}{\mbox{\thinspace g\thinspace s$^{-1}$}}
\newcommand{\IVl}{\mbox{\thinspace{\sc iv}}$\,\lambda\,$}
\newcommand{\IIl}{\mbox{\thinspace{\sc ii}}$\,\lambda\,$}
\newcommand{\IIIl}{\mbox{\thinspace{\sc iii}}$\,\lambda\,$}

\newcommand{\Vl}{\mbox{\thinspace{\sc v}}$\,\lambda\,$}
\newcommand{\VIl}{\mbox{\thinspace{\sc vi}}$\,\lambda\,$}
\newcommand{\mdot}[1]{\mbox{$\dot{M}_{\rm #1}$}}
\newcommand{\rwd}{\mbox{$R_{\rm WD}$}}
\newcommand{\twd}{\mbox{$T_{\rm WD}$}}

\newcommand{\EXPU}[3]{\mbox{\rm $#1 \times 10^{#2} \rm\:#3$}}  
\newcommand{\POW}[2]{\mbox{$\rm10^{#1}\rm\:#2$}}
\newcommand{\CHINU}{\mbox{$\chi_{\nu}^2$}}
\newcommand{\fuse}{\mbox{{\it FUSE}}}
\newcommand{\iue}{\mbox{{\it IUE}}}
\newcommand{\orfeus}{{\mbox{\it ORFEUS}}}
\newcommand{\vsini}{\mbox{$v\:\sin{(i)}$}}
\def\VEL{\:{\rm km\:s^{-1}}}
\newcommand{\MSOL}{\mbox{$\:M_{\sun}$}}

\shorttitle{The Far Ultraviolet Spectrum of Z Cam in Quiescence}
\shortauthors{Hartley et al.}

\received{2003 June 26}
\begin{document}

\title{The Far Ultraviolet Spectrum of Z Cam in Quiescence and Standstill\footnote{Based on observations made with the NASA-CNES-CSA
{\it Far Ultraviolet Spectroscopic Explorer} (\fuse\/). \fuse\/ is
operated for NASA by Johns Hopkins University under NASA contract NAS
5-32985.}}
\author{Louise E. Hartley\footnote{Current Address: Centre for
Astrophysics and Supercomputing, Mail 31, Swinburne University of
Technology, Hawthorne, Vic 3122, Australia} \  \& Knox S. Long}
\email{lhartley@astro.swin.edu.au, long@stsci.edu}
\affil{Space Telescope Science Institute\\3700 San Martin Drive\\
  Baltimore, MD 21218}
\author{Cynthia S. Froning}
\email{cfroning@casa.colorado.edu}
\affil{Center for Astrophysics and Space Astronomy, \\ University of Colorado, 593 UCB, Boulder, CO 80309}
\author{Janet E. Drew}
\email{jdrew@imperial.ac.uk}
\affil{Imperial College of Science, Technology and Medicine\\Blackett
Laboratory, Prince Consort Road\\London, SW7 2BW}

\begin{abstract}
We have obtained {\it Far Ultraviolet Spectroscopic Explorer}
(905--1187\AA) spectra of the non-magnetic cataclysmic variable
Z~Cam during a period of quiescence. The spectrum resembles that
of a hot (57,000\,K) metal-enriched white dwarf. A high effective
temperature is consistent with the expectation that dwarf nova
systems that show standstills, as Z Cam does, have higher than
normal time-averaged mass accretion rates. It is also consistent
with current estimates for the mass and distance to Z Cam.
A white dwarf model in which 29\% of the surface has a temperature
of 72,000 K and the rest of the surface is at 26,000\,K also
reproduces the spectral shape and the continuum flux level at the
nominal distance and WD mass, and is a somewhat better statistical
fit to the data. A second component could be due to a rotating
accretion belt, a remnant of an outburst that ended eleven days
prior to the \fuse\ observation, or ongoing accretion. We favor
the uniform temperature model for Z Cam, largely because the data
do not require anything more complicated. We have compared the
quiescent spectrum with an archival spectrum of Z~Cam in
standstill, obtained with \orfeus. The standstill spectrum is
described well by a disk accreting at
\mdot{disk}=\EXPU{7_{-2}^{+4}}{16}{\gs}, where the errors depend
on the assumed inclination.

The quiescent spectra cover a full orbital period and are
time-resolved to $\sim500\,$s. No variability was observed in the
continuum during the observation, but the depth of many of the
absorption lines increased between orbital phases 0.65 and 0.81.
We attribute this effect to absorption by material associated with
the accretion stream, which is easier to understand if the
inclination is near the maximum allowed value of 68\degr.
\end{abstract}


\section{Introduction}
\label{s:intro} Cataclysmic variables (CVs) are interacting binary
systems in which a white dwarf (WD) accretes matter from a
low-mass companion. In non-magnetic CVs, the in-falling matter
forms a disk around the WD. Dwarf novae (DNe) are disk-accreting
CVs that undergo sharp rises in luminosity, during which the
optical brightness typically increases by 2 to 5\,mag. The
outbursts occur when a sudden increase in the gas viscosity (due
to a change in the ionization state of the gas) allows matter to
flow more rapidly through the disk, causing an increase in the
accretion luminosity.

In most DNe, outbursts last for a day to a few weeks. In the
subclass of DNe known as U Gem systems, outbursts end with a
steady decline to minimum that has a typical timescale of a day.
However, in Z~Cam systems, the decline from outburst occasionally
stalls about 1\,mag below the peak. Such `standstills' can last
for days to, in a few cases, years. Other non-magnetic CVs
(nova-like variables) appear to remain persistently in the
outburst state. The Z~Cam standstill state is comparable to that
of a nova-like variable, i.e. a steady, high state, but somewhat
lower in luminosity than the peak of a dwarf nova outburst.

The difference in outburst behavior between U~Gem, Z~Cam and
nova-like systems is thought to reflect a difference in the rate
of mass flow from the secondary (\mdot{2}). In DNe, \mdot{2} is
below the upper stability boundary (\mdot{crit}) for which DN
outbursts can occur, whereas in nova-like systems
$\mdot{2}>\mdot{crit}$. In Z Cam-type stars $\mdot{2}$ lies very
near, but below $\mdot{crit}$ most of the time. Standstill phases
are triggered when an increase in $\mdot{2}$ \citep{98king} is
accompanied by a decrease in $\mdot{crit}$ as a response to
heating of the outer edge of the accretion disk by the increased
mass transfer \citep{01buat,01stehle}. Thus, $\mdot{2}$ becomes
greater than $\mdot{crit}$ and the disk assumes a stable high
state (like a nova-like variable). It has been difficult to
measure mass transfer rates in CVs observationally, and as a
result reliable empirical determinations of \mdot{2} and
\mdot{crit} are lacking. On theoretical grounds, \mdot{crit} is
thought to be about \EXPU{3}{17}{\gs} when heating of the mass
transfer stream is taken into account \citep{02schreiber}.

Z~Cam is the brightest (in outburst) and most extensively studied
object of its class. It has an orbital period $P_{\rm
orb}=0.289\,$d, which implies that it, like other standstill
systems, is above the period gap. (A complete list of system
parameters is presented in Table~\ref{t:param}.) The difference in
the magnitude at outburst maximum ($m_v$=10.4) and quiescence
($~$13) is small for DNe in general, but typical of standstill
systems. The standstill $m_v$ averages 11.5 \citep{98oppenheimer}.
Far ultraviolet (FUV) spectra of Z~Cam in outburst were obtained
with the Hopkins Ultraviolet Telescope (HUT)
\citep{91long,93long,97knigge}. The outburst spectrum resembles
that of a steady-state accretion disk with a mass accretion rate
\mdot{disk}=\EXPU{3}{17}{\gs}, modified by resonant scattering by
material in a wind \citep{97knigge}.

Here, we attempt to complete the picture of Z Cam by describing
FUV spectra of Z Cam in the quiescent and standstill states. We
utilize spectra of Z Cam in quiescence that we have obtained using
{\it Far Ultraviolet Spectroscopic Explorer} (\fuse), and a
spectrum of Z~Cam in standstill obtained with \orfeus\ telescope
that we retrieved from the MAST archive. Our discussion of the
effort to understand the two datasets is organized as follows: In
Section~\ref{s:obs}, we describe the \fuse\ data reduction and the
mean spectrum of Z Cam in quiescence. In Section~\ref{s:cont}, we
attempt to fit the spectrum of Z Cam in quiescence, while in
Section~\ref{s:var} we analyze the time variability of the
spectral lines.  In Section~\ref{s:stand}, we describe the
\orfeus\ spectrum of Z Cam in standstill as well as our attempts
to model that spectrum. Finally, in Section~\ref{s:discus}, we
evaluate the available spectral models and compare the properties
of Z~Cam in its different luminosity states.

\section{Observations and Qualitative Description of the \fuse\/ Data}
\label{s:obs}

\fuse\ provides a resolving power of $\sim$20,000, in the
905--1187\,\AA\ range \citep{00moos,00sahnow}. Z~Cam was observed
with \fuse\ on 2002 February 9, from 13:19:54~UT to 21:06:12~UT.
As indicated in Fig.\ \ref{f:aavso}, the observations took place
about eleven days after the return to quiescence from a short
outburst and five days before the next outburst. The observations
were made through the LWRS ($30\arcsec\times30\arcsec$) aperture.
The data were obtained in the histogram mode, because Z Cam can
(in outburst) exceed the brightness limit for the
high-time-resolution mode.  A total of 24,436~s of useful data was
obtained. This comprised 43 separate exposures, ranging in length
from 534\,s to 584\,s.

The raw data were extracted and calibrated with version 2.2 of the
CALFUSE pipeline and version 12 of the CALFUSE calibration files.
The spectrum from each \fuse\ channel was examined separately to
ensure that the flux calibration was consistent enough for the
spectra to be spliced together. Some data at the edges of each
detector segment where the flux calibration is often uncertain
were masked out, as was the `worm' feature that is present in the
LiF1b channel. The spectra from the SiC1a channel indicated
anomalously higher fluxes in  the range 1000-1065 \AA\, and so
this region was excluded. The remaining data from the four
channels were combined to create spectra at 0.1\AA\ binning. Our
process for combining the data consists of summing all the data
points that fall within a given wavelength bin, weighting each
according to the detector pixel sensitivity.

A time-averaged spectrum was created by summing the individual
spectra, weighted according to exposure time. This is shown in
Fig.\ \ref{f:spectrum}. The interstellar lines have measured
widths of $\sim$0.1\,\AA\ (about the expected spectral
resolution), indicating that the four channels remained
well-aligned over the course of the observation. This is supported
by the flux levels of the overlap regions between spectral
channels.

The quiescent spectrum has a fairly flat continuum with peak flux
lying near 1000\,\AA, at a flux level of
$2.2\times10^{-13}$\funits. The continuum flux falls by only
$\sim25\,$\% from its peak level before Lyman line blanketing sets
in at 912\AA.
Table~\ref{t:lines} lists the spectral lines observed in the
\fuse\/ spectrum of Z~Cam. The equivalent width (EW) measurements
are made from the time-averaged spectrum, except for those lines
that are seen to be obscured by airglow in Fig.\ \ref{f:spectrum};
in this case the measurements are made from a mean spectrum
constructed from exposures that were recorded during spacecraft
night (as indicated in the image header). The night-time exposures
make up about 30\% of the total observing time and are relatively
free of distortion by airglow lines. Of absorption lines intrinsic
to Z~Cam, we see the Lyman series up to at least $n=6$, and a
range of metals, including C\II--\III, N\II--\IV, O\VI,
S\III--\VI\ and Si\III--\IV. The absorption lines are smooth and
are fit well by Gaussian profiles with FWHM of $<1000$\kmsd.  Some
of the absorption lines cut into broad emission profiles. These
include the lower ionization lines of C\IIIl977 and C\IIIl1176,
S\IV$\,\lambda\lambda$1062,1073 and \mbox{Si\IIIl1108--1113} and
the higher ionization O\V$\,\lambda\lambda$1032,1038 lines. None
of the absorption lines fall to zero flux at the line core: the
deepest absorption lines are O\VI\ and C\III, which have
absorption minimum flux levels of, respectively, 30 and 40\%\ of
the continuum.

A careful inspection of the mean spectrum indicates that the flux
below the Lyman limit is not zero, but averages
\EXPU{4}{-14}\funits. Judging from earlier FUV observations
(Knigge et al. 1997, also Section~\ref{s:stand}) and given the
hydrogen column density ($N_H=$\EXPU{3}{19}{cm^{-2}}) to Z Cam
\citep[Mauche 1995, private communication as quoted
by][]{97knigge}, we expect to see no flux from Z Cam in this
region. Calibration of \fuse\/ spectra involves a background
subtraction step and for histogram data a time-averaged background
is used. If the average background during the observation were
higher, then the extra background will appear as extra flux after
the \fuse\/ pipeline processing. The background is a combination
of scattered light and internal detector background and does not
follow the effective area curve of \fuse\/ for valid photons. As a
result, the effect of poor background subtraction is most severe
when the effective area is smallest (or at the short wavelength
end of the detector).

The scattered-light portion of the background is much higher in the
spacecraft daytime.  Therefore, we checked the
mean spectrum from the full observation against one constructed from
night-time only exposures.  For the night-time spectrum, the flux below
the Lyman limit averages \EXPU{1}{-14}\funits (one quarter that of the
mean spectrum). At wavelengths longer than 915\AA\ the
discrepancy is significantly smaller, with the night-time continuum
being on average $\sim5\,$\% less than in the mean spectrum, although
in the two spectra the depth of the interstellar Lyman lines differs
by about the same amount as the flux level below 915\AA. At the minima
of absorption lines intrinsic to Z~Cam (excluding those filled in
by airglow in the mean spectrum) the day-time flux is, at worst 15\%
less than the night-time flux level, with the effect being greatest in lines lying below 1000\AA.

Given that the flux offsets are small we are confident that the
background subtraction problem does not affect the spectrum
significantly at most wavelengths and does not prevent an accurate
qualitative assessment of the spectrum.  However, in applying spectral
model fits (in Section~\ref{s:cont}) we use only the night-time
data. This is primarily because the absence of many of the strong
airglow lines that are seen in the mean spectrum allows more data
points to be included in the model fits, but also because we are more
confident that the flux level of the night-time only data is closer to
the true observed flux.

\section{Analysis of the Quiescent Continuum}
\label{s:cont}
A comparison of model flux levels with the observed fluxes calls for a
reliable measurement of the distance to Z Cam. This has been estimated
variously at $>190\,$pc \citep{85berriman} and $<390\,$pc
\citep{80kiplinger}, based on the fraction of infrared and optical
luminosity that can be attributed to the secondary star. A recent
parallax measurement, combined with proper motions and absolute
magnitude constraints, puts the distance somewhat closer at
$163^{+68}_{-38}\,$pc \citep{03thorstensen}; we shall use this last
distance in this paper.

\subsection{White Dwarf Model Fits}
\label{ss:wdmods}

WD model spectra were generated using TLUSTY and SYNSPEC software
\citep{88hubeny,94hubeny,95hubeny}.  The model spectra covered
effective temperatures from 13,000\,K to 100,000\,K and rotation
rates of 0 to 1200\kmsd. We adopted $N_{\rm
H}=3.0\times10^{19}\,$cm$^{-2}$ \citep[Mauche 1995, private
communication as quoted by][]{97knigge}, E(B-V)$=$0.0, and $\log
g=8.5$ (strictly $\log{g}=8.6$ is expected for a 1$M_\odot$ WD).
The parameters of all the model fits are listed in
Table~\ref{t:fits}.

We first fit non-rotating DA models to the mean night-time
spectrum, after masking all of spectral lines other than those
associated with hydrogen. The models were fit by $\chi^2$
minimization. The best-fit DA model has $\twd =$51,200\,K. As
shown in Fig.\ \ref{f:dafit}, the model is a reasonably good match
to the continuum slope and the peak of the continuum at $\sim$1000
\AA. The best-fitting DA model falls below the data from the Lyman
limit to about 950 \AA; at the limit the model flux lies at about
$\sim60\%$ of the observed flux. Varying the interstellar $N_{\rm
H}$ about its modest value has a negligible effect on the
appearance of the best-fit model spectrum. Increasing the
interstellar reddening from zero to E(B-V) $=0.03$ \citep[the
upper limit given by][]{87verbunt} increases the best-fit
temperature marginally, and also lowers the model peak flux level
slightly, to better match the observed level.

Assuming the entire WD is visible, the normalization of a model is
a measure of the solid angle of the WD and can be combined with
the distance to give the WD radius. For the best-fit DA model we
arrive at
\mbox{$\rwd=$\EXPU{4.7^{+0.8}_{-0.3}}{8}{cm}}, where the errors
correspond to the upper and lower distance limits. Adopting the WD
mass-radius relation of \citet{88anderson} and $M_{\rm
WD}=0.99\pm0.15\,M_\odot$ \citep[][quoted in
\citealt{97knigge}]{83shafter}, the expected \rwd\ for Z~Cam is
\EXPU{5.8^{+1.1}_{-1.2}}{8}{cm}. The value \rwd\ (and hence the
mass) determined from the fit is therefore consistent with the
value of \rwd\ predicted from Shafter's mass estimate and the
normal WD mass-radius relation.

The DA model spectrum does not fit the higher order Lyman lines
particularly well. They are broader and deeper in the model
spectrum than the observed lines. Increasing \twd, or decreasing
the surface gravity of the model narrows the predicted profiles,
but not enough to bring the models and the data into good
agreement. Changing these parameters also causes the spectral
slope to become steeper, worsening the model fit to the continuum.
One possibility is that there is an emission component, possibly
from the disk, that contributes to the narrowness of the observed
lines.  The H\I\ emission is illustrated most effectively by the
difference spectrum in Fig.\ \ref{f:dafit}. The emission
components around other lines (as noted in Section~\ref{s:obs})
can also be made out in the difference spectrum, although they
appear to be weaker than the H\I\ emission.  We contrast this to
the result of \citet{98gaensicke}, who find that the lack of
strong Lyman absorption in the UV spectrum of AM~Her in the high
state can be explained by irradiation of the WD surface by a hot
spot that radiates as a blackbody, {\em without} noticeable
emission or absorption features.

We next considered WD with metal-enriched photospheres.  Solar
abundance models were created for a range of temperatures from
13,000\,K to 100,000\,K and WD rotational velocities from 0 to
1200\kmsd. Before fitting the spectrum, the
O\VIl$\lambda\,$1031,1039, S\VIl$\lambda\,$933,944 and N\IVl924
lines were masked out, as WD surface temperatures are not
typically high enough to form these high-ionization lines.

The best-fit solar-abundance WD model is shown in Fig.\
\ref{f:wdfit1}. It has $\rwd=$ \EXPU{4.4}
{8}{cm}, $\twd =57,200\,$K and $\vsini=330$\kmsd. The peak in the
continuum at 1000\,\AA\ and the slope of the continuum are a good
match to the data. However, like the DA model the solar abundance
model underestimates the flux at the long- and short-wavelength
ends of the spectrum.  The effect is slightly less in the solar
abundance case, because the hotter WD does a better job
reproducing the short-wavelength emission. As was the case with
the DA model fits, the best fit parameters indicate that the WD is
at somewhat greater distance than 163 pc and/or has some what
greater mass than 0.99\MSOL.  However, given the error bars, the
various estimates are self-consistent.

The solar abundance WD model does a good job of reproducing some of
the stronger absorption lines, e.g. C\IIIl977, N\IIl1085,
Si\IVl$\lambda\,$1122,1128 and C\IIIl1176. In the model the N\IVl924,
N\IIIl989, S\IIIl1012-1021 and S\IVl$\lambda\,$1062,1073, S\IIIl1077
and Si\IIIl1108-1113 lines are significantly weaker then in the data,
whereas the Si\IVl1066 line is much stronger.

In other quiescent CV, two-temperature (2T) WD model fits have
been known to give a significant improvement over the
single-temperature models
\citep[e.g.][]{93long,96gaensicke,02szkody}. In non-magnetic CVs,
the second component is attributed to matter arriving onto the
surface of the WD from the accretion disk. The accreted matter is
hotter than the WD and travels at the Keplerian rotational
velocity, and so the high-temperature, rapidly-rotating gas can
form an `accretion belt' on the WD surface \citep{78kip}.

To see whether model fits of this type provide a better
representation of the spectrum, we fit the data to models
consisting of two different temperatures, allowing the
normalization and rotational velocity of each to vary
independently.  The best fit has one component with temperature
$T_1$ of 26,300 K covering 71\% and a second component with $T_2$
of 71,700 K covering 29\% of the surface.  Statistically, the more
complex model with \CHINU\ of 9.4 is better than the
single-temperature fit with \CHINU\ of 10.1, but qualitatively
there is little difference between the fits and neither is
``statistically acceptable'' in the sense that \CHINU\ is still
considerably greater than one. Therefore physical arguments will
ultimately guide the interpretation.)  For the 2T fit, the
dominant contribution to the emission is due to the higher
temperature component.  At 1100 \AA, 78\% of the flux is from the
high temperature component and shortward of 1000 \AA\ essentially
all of the flux is from the hot component. The 2T fit does result
in a larger radius of \EXPU{6.0}{8}{cm} for the WD, which is
typically the case when 1T and 2T models are compared. A value of
\EXPU{6.0}{8}{cm} (at the nominal distance for Z Cam) is close to
the value of \EXPU{5.8^{+1.1}_{-1.2}}{8}{cm} predicted from the
estimated mass.

We have performed tests to see how the temperature and
contribution of the hot component ($T_2$) varies as a function of
the temperature of the low temperature ($T_1$) is adjusted. For
$T_1$ between 23,000 and 37,500 K, it is possible to fit a
two-component WD model with a radius in the range allowed by the
error limits on the WD mass (Table~\ref{t:param}). As the
temperature of the low-T component increases, the fraction of the
surface covered by the hot component also increases. At $T_1$ =
23,000\,K, the hot component has $T_2$=70,000\,K and covers 27\%
of the surface, whereas at 37,500\,K the hot component has
$T_2$=75,000\,K and covers 41\,\% of the surface. Statistically,
there is little difference between the goodness of these 2T fits.
They all have \CHINU\ in the range 9.4 to 9.9.

The values of the rotation velocity $\vsini$ for our best-fit 2T
model are 170 and 320 $\VEL$ for the low and high temperature
components, respectively. There is no real (positive) evidence
that the second component, if it exists, is rotating rapidly, and
hence there is no ``smoking gun'' for a rapidly rotating accretion
belt. This is not entirely surprising though, given the fraction
of the flux contributed by the high temperature component.

Z~Cam was extensively observed with the {\it International
Ultraviolet Explorer} (\iue\/) between 1979 and 1991, although
only twice during quiescence. The flux level at the long
wavelength end of the \fuse\/ spectrum provides an excellent
overlap with the short-wavelength end of the Z~Cam quiescent
\iue\/ spectrum SWP18844 \citep[as presented in][]{86szkody} and
SWP37155.  We rebinned the \fuse\/ spectrum to the 1.7 \AA\ to
match the wavelength spacing of the LOWRES \iue\/ dataset and
merged it with SWP18844. We then created model spectra to match
the coarser spectra resolution of \iue.  At this resolution, there
is little sensitivity to $\vsini$ and so all of the models were
constructed with $\vsini$ of 300 $\VEL$. A solar abundance model
fit to the combined \fuse\/ and \iue\/ spectrum (Fig.\
\ref{f:dafuseiue}) has $\twd =46,600\,$K and
\rwd=\EXPU{6.0}{8}{cm}. At $\lambda\ga1500\,$\AA\ the model flux
level is somewhat lower than is observed and at around 1000\,\AA\
the model continuum level lies slightly above that of the observed
spectrum.  The larger value of \rwd\ is a direct result of the
lower temperature in these fits.

In Fig.\ \ref{f:dafuseiue}, we also plot the best 2T WD fit to the
combined spectrum.  In this case, $T_1$ is 17,600 K and this
component covers 97\% of the surface, while $T_2$ is 82,000K and
covers only 3\% of the surface.  In terms of flux, the low T
component contributes 27\% at 1100 \AA\ and 46\% at 1500 \AA.
There is a significant improvement in the statistical quality of
the fit from \CHINU\ of 8.2 to 4.9.  The 1T model has difficulties
at the short and long wavelength ends of the spectrum; the 2T
model (predictably) addresses these difficulties since the high T
component provides flux near the Lyman limit while the low T
component provides flux at the long wavelength portion of the
spectrum. However, \rwd\ is now larger than desired.  If one
constrains the normalization to place the system at the nominal
distance and radius ranges, tests show that larger $T_1$ and lower
$T_2$ are required.  As was the case for when only the \fuse\ data
were fitted, there is a locus of values for $T_1$ and $T_2$ that
provide qualitatively similar fits to the data.

\subsection{Accretion Disk Model Fits}
\label{ss:diskfits}

Given that Z Cam was observed in the quiescent state, we do not
expect that steady-state accretion disk models would provide a
good description of the \fuse\ spectra.  Nevertheless, it is
important to check that this is not the case. Therefore, we
constructed model disk spectra from summed, area-weighted,
Doppler-broadened spectra of stellar atmospheres of the
appropriate temperature and gravity for each disk annulus. As for
the WD models, we adopted $E(B-V)=0$ and $N_{\rm
H}=3.0\times10^{19}\,$cm$^{-2}$.  Disk spectra were created for a
range of \mdot{disk}\ from $10^{14}$ to \POW{19}{g~s^{-1}}, and
normalized to a distance of $163\,$pc. Since the inclination for Z
Cam is not tightly constrained ($57\pm11\degr$), we created model
spectra for inclinations of 46\degr, 57\degr, and 68\degr.  We
then fit the data using the same wavelength intervals as the solar
abundance WD fits.  Unless otherwise indicated, these fits have
one variable \mdot{disk}, since in most cases we did not allow the
normalization (or alternatively the distance) to be a free
parameter.

Fig.\ \ref{f:diskfit1} shows the best-fit disk spectrum with
\mdot{disk}=\EXPU{9.1}{15}{\gs}. This model matches the slope and
level of the continuum between 1050 and 1185\AA, but bluewards of
\lyb\ the model flux declines steeply to zero. As expected from a
Keplerian disk, the rotational velocities are too high to fit the
narrow absorption lines.  Since we are not allowing the
normalization to vary in these fits, the exact values of
\mdot{disk} that we derive are dependent both on the distance to Z
Cam and on the inclination.  The inclination of Z Cam cannot be
very much larger than 68\degr\, or eclipses would be observed.  If
we assume this inclination, then we obtain \mdot{disk} of
\EXPU{1.3}{16}{\gs} and a slightly better fit in terms of \CHINU.
This apparent improvement in the fit is due to the fact that
higher \mdot{disk} model has a flatter continuum slope at shorter
wavelengths.  This said, the fits are considerably worse than both
the uniform T and two-T WD models.

Better fits would be obtained if the normalization of the spectra
were not fixed.  Indeed the best fit to the spectral shape would
suggest \mdot{disk} of \EXPU{1.8}{17}{\gs}, a physically
implausible value for a quiescent system, and an equally
implausible distance of 710 pc.

We also combined an accretion disk component with the WD models.
We fit two-component models wherein the first component was a WD
with $\log g=8.5$ and the second component was an accretion disk
of $i=57\degr$ (the resultant spectrum was essentially unchanged
when the disk inclination was varied to 68\degr).   In the
combined disk and WD model the WD is at 62,000\,K, with
$\rwd=\EXPU{4.0}{8}{cm}$ and the disk has
\mdot{disk}=\EXPU{1.7}{15}{\gs}. The resultant model differs
negligibly from the WD model fits (see Fig.\ \ref{f:wdfit1});
\CHINU\ is 9.8, slightly worse than produced in the 2T WD fits
(9.4). The WD component dominates the model spectrum, while the
disk component acts basically only to flatten the continuum slope,
by providing a small flux contribution at long-wavelengths. Since
in quiescence, one does not necessarily expect the spectrum to
mimic that produced by steady state disk models, we also
experimented with models in which the disk contribution was
assumed to be a power law ($f_{\nu}\propto \nu^{-\alpha}$).  These
models produce similar fits statistically (\CHINU=8.7) to fits
using a  WD and a steady state disk built from stellar spectra.
The power law contributes about 33\% of the flux at 1100 \AA\ and
has a slope $\alpha$ of 1.6.   The WD contributes the bulk of the
flux, and has a temperature of 44,900 K, \rwd\ of
\EXPU{4.3}{8}{cm}, and \vsini\ of 210$\VEL$.

For completeness, we carried out the same disk model analysis for
the combined \iue\ and \fuse\ data.  The results are almost
identical to that of the \fuse\ data alone. A careful inspection
of Fig.\ \ref{f:fuse_iue_disk}, where the results are shown,
indicates that the main observational disagreement between disk
and WD models for Z Cam is in the \fuse\ wavelength range.  We
note in passing that \CHINU\ for the combined WD and disk fit (5.0
for 57\degr) is better than for simple 1T WD or disk models, and
almost identical to that obtained  for 2T WD models (4.9). The
inferred WD radius for the combined WD and disk model fit suggests
a very massive WD, in excess of 1.35\MSOL\ and much greater than
estimated by \cite{83shafter}. One way to lower the mass would be
to allow the distance to be larger. But much more probably, the
result is an artifact of our limited knowledge of the spectrum of
a quiescent disk, which we, in the absence of any fully-calculated
alternative, have represented as that expected from a steady-state
optically-thick disk.

\subsection{The Quiescent Line Emission}
It has been suggested that in quiescent CV the accretion disk is
truncated at its inner edge, due to irradiation by the white dwarf
\citep[e.g.][]{97king}. If we assume that the emission components seen
around some of the absorption lines are formed in a Keplerian disk,
and hence are Doppler broadened by the disk rotation, we can estimate
the smallest radius at which the emission is formed.

Using the Starlink software {\sc dipso}, we fit the
S\IV$\,\lambda\lambda\,$1063,1072, Si\IIIl1108--1113 and
C\IIIl1175 lines after first normalizing the continuum flux. We
fit a Gaussian absorption and emission line to each component of
the multiplet (except in the case of C\IIIl1175, which is fit as a
single line). These are shown in Fig.\ \ref{f:emission}. After
subtracting the multiplet splitting, the emission components all
have widths of about $\sim$4000\kms\ (FWZI). This translates to a
rotational velocity of 2400\kmsd, or a Keplerian disk radius of
\EXPU{2.3}{9}{cm}, which is $\sim$4\rwd\ \citep[][for example,
predicts that the quiescent disk truncation radius is
$\sim(1-5)\rwd$]{97king}. This suggests that the emission lines
could be formed by material rotating with the quiescent accretion
disk.

At present we have only a poor idea of exactly what observational
signatures from a cold optically-thin disk would be detectable in
the FUV. However, there is substantive evidence that there is an
additional component to the quiescent continuum and some attempts
have been made to model the structure of this as a coronal layer
above the quiescent disk \citep[e.g.][]{94meyer,95liu}. There
currently exist no quantitative UV spectral models for such a
corona, so in order to get some idea of what these properties
might be, we constructed TLUSTY models of an optically thin plasma
veil, at various temperatures and densities, which were then
rotated as a solid body, to mimic the disk rotation. We then took
a simple optically thin emission prescription for the emitted flux
($F_\nu=B_\nu(1-e^{-\tau_\nu})$). The optically thin plasma was
able to reproduce the relative strengths of the emission
components seen around some absorption lines. However, there was
some degeneracy between various H column densities, electron
densities and temperatures. Hence,  we found it impossible to
estimate, from these models, a possible physical structure for the
corona.

\section{Variability on the orbital timescale}
\label{s:var} The 43 histogram exposures cover just over one
orbital period of Z~Cam from 0.94 to 0.06, with a phase
uncertainty of 0.015 (using the spectroscopic ephemeris of
Thorstensen \& Ringwald, 1995). The phase coverage is continuous,
apart from gaps at phases 0.57--0.63 and 0.81--0.88. We are thus
confident that the mean spectrum is representative of the average
over the full orbital period, rather than a reflection of some
orbital phase-dependent characteristic.

We created a continuum light curve by summing the flux over
selected spectral windows. For the light curve we used only data
from the LiF1 channel. \fuse\/ is guided on this channel and so it
should be less susceptible than the others to channel
misalignment. To measure flux variation across the entire LiF1
range, we first excluded regions contaminated by airglow and the
worm region (see Section~\ref{s:obs}) and summed the flux over the
ranges 996--1024, 1029--1038, 1046--1081, 1095--1103 and
1170--1185\AA. Continuum flux varied by up to $\sim15\,$\% either
side of the mean. However, given the noted discrepancy between the
day and the night-time flux levels (see Section~\ref{s:obs}), we
found no evidence of variability in the continuum flux level in
excess of that which could be associated with the periodic
transition of the spacecraft from day to night time.

The spectroscopic ephemeris of Thorstensen \& Ringwald (1995) is
accurate enough to radial-velocity correct each exposure onto the
WD rest frame, with errors of about 25\kmsd. The spectra presented
in this section are plotted after this correction has been
applied. Fig.\ \ref{f:variation1} shows the 1005--1045\,\AA\
region plotted in phase bins of 0.1$\,P_{\rm orb}$.  The
1100--1133\AA\ spectral range encompasses the C\IIl1010,
S\IIIl1012--1021, \lyb\ and O\VIl$\lambda\,$1032,1037 lines and
gives a good representative sample of the line variability seen
across the spectrum.

In the time-averaged spectrum the flux minima of the absorption
lines are all centered at between 100 and 200\kms\ redwards of the
transition rest velocity. Throughout the observation there is
almost no shift in the position of this feature, except for during
phase 0.5 to 0.7 where the entire line profile shifts bluewards by
about 100--200\kms\ (i.e. back to closer agreement with the rest
velocity, see Fig.\ \ref{f:variation1}). That the mean line
profiles are centered redwards of the rest-velocity could well be
an artifact of the radial velocity correction. We take the
zero-phase ephemeris from the inferior conjunction of red star
emission lines, rather than using the H$\alpha$ emission line
ephemeris, as the latter may be formed in an extended outflow
\citep[see][]{95thorstensen}. This therefore, does not take into
account gravitational redshifting, which is 76\kms if $M_{\rm
WD}=0.99$ (or 108\kms if $M_{\rm WD}=1.14$). If the absorption
lines are formed on the surface of the WD, then this accounts for
a portion of the redshift of the mean line profile centroids.

There is little variability in the strength and width of the
absorption lines, except between phases 0.65 and 0.81 (Fig.\
\ref{f:addabs}). Then there is a sharp increase in the strength of
most of the absorption lines and we clearly see some lines that
are barely apparent in the mean spectrum, e.g.
P\VIl$\lambda\,$1118,1128 and the Si\IIIl994--997 triplet. At the
same time there is no sign of an increase or decrease in continuum
flux. The FWHM of the lines increases by no more than a few tens
of \kmsd. The additional absorption mainly affects the depth of
the line absorption. There is no shift bluewards or redwards from
the mean line profile.

The increased absorption is seen across the full range of atomic
species and ionization stages, for example S\IIIl1012--1021,
S\IVl$\lambda\,$1062,1073 and S\VIl$\lambda\,$933,944. The higher
ionization S\IV\ and S\VI\ lines increase in EW by around 30\,\%.
On average, the increase in line EW at phase 0.65--0.81 is also
$\sim30$\,\%. The S\III\ triplet shows increased absorption by up
to 300\,\%\ in the mean EW of the S\IIIl1013 component, but only
40\,\%\ in the S\IIIl1021 component.   Some lines, such as
Si\IIIl1108--1113, show a smaller variation ($\sim10\,$\%);
others, such as Si\IIIl997 and P\Vl1118, increase their EW by two
or three times; and others, such as Si\IIIl$\lambda\,$994,995,
appear only during the period of enhanced absorption.  There are a
few lines, most notably N\IVl924, O\VIl$\lambda\,$1032,1038, and
C\IIIl977 and 1176 that do not increase in EW at phase 0.65--0.81.
As the lines that show increased absorption cover a range of
excitation energies we can rule out any obvious
temperature-dependence. An increase in the absorption line
strength could stem either from an increase in the column density
of absorbing ions, or from an increase in their covering factor
against the continuum background (or a combination of the two).
Given that, at the time of the increased absorption, we see no
evidence of an accompanying alteration of the ionization balance,
it seems most likely that we are seeing a shift of material into
the line of sight and/or an increase in the continuum covering
fraction of absorbing ions. We are unable to say whether the
narrow absorption endures beyond phase 0.81 as there is a gap in
the data from phase 0.81 to 0.88. However, when the coverage picks
up again at phase 0.88, the additional absorption has disappeared.

\section{The FUV spectrum of Z~Cam in standstill}

\label{s:stand} Z~Cam entered standstill on 1992 September 23. Z
Cam was observed with the Berkeley Extreme and Far-UV Spectrometer
(BEFS), on board the \orfeus\ telescope \citep{98hurwitz}, on 1993
September 18. A single 1477s spectrum was obtained. The system
remained in standstill for a further 100 days, before declining to
quiescence \citep[see the lightcurve in][]{98oppenheimer}. During
the standstill the system brightness can fluctuate by about
0.3\,mag either side of the average magnitude of 11.5, but the
timing of this observation so far into the standstill assures us
that \orfeus\ captured Z~Cam in a true standstill phase.

We retrieved the BEFS spectrum, pre-calibrated, from the MAST
archive. The spectrum covers the range 700--1175\AA\ at a spectral
resolution of about 0.2\AA\ in the two longer-wavelength channels
of the spectrometer. However, no flux was detected below 912\AA,
so we have used only data from the longest-wavelength channel
(900--1175\AA).

In Fig.\ \ref{f:standstill} we have plotted the BEFS and the
\fuse\ spectra together. Despite the order-of-magnitude difference
in continuum flux, the two spectral shapes are very similar. The
standstill spectrum peaks at $\sim1060\,$\AA, at
\EXPU{\sim2}{-12}{\funits}. This is to the red of the peak in the
quiescent spectrum (at $\sim1000\,$\AA). On the longer-wavelength
side of the peak, both have similar continuum slopes. Both spectra
show significant flux down to the Lyman limit, although the
continuum between 912\AA\, and the peak flux is slightly flatter
in the quiescent spectrum than in standstill.

Most of the absorption lines seen in the standstill spectrum as we
are also seen in quiescence (see Table~\ref{t:standstill}). In the
BEFS spectrum the absorption equivalent widths are generally
30-60\% greater in the higher-ionization lines (e.g. O\IV, S\IV\
and S\VI) than in the \fuse\ spectrum, whereas the
lower-ionization lines (e.g. Si\III\ and S\III) are weaker by a
factor of at least 2. Note also the appearance of the
P\Vl$\lambda\,$1118,1128 lines at standstill (which are only seen
at phase 0.65--0.81 in quiescence). This is indicative of a
general move upward in the ionization state between quiescence and
standstill. The standstill line widths (FWHM) are generally about
100--200\kms\ narrower than the quiescent lines, with the
exception of C\III, N\III\ and O\VI. In standstill, there is no
sign of the broad line emission that is seen in quiescence. Nor
for that matter are there indications of the wind that is observed
in outburst in Z Cam, e.g. P-Cygni-like profiles or even
blue-shifted absorption features.

We repeated the model-fitting exercise that was carried out on the
quiescent spectrum, beginning with solar abundance WD models (as
for Section~\ref{s:cont}). In order not to bias a comparison to
the \fuse\ data, we excluded the same regions of the spectra as
were used for the \fuse\ analysis. In general, the best-fit values
of \CHINU\ were lower for the \orfeus\ than the \fuse\ data, a
fact that is directly attributable to the statistical quality of
the two datasets.

The best-fit WD model is shown in Fig.\ \ref{f:standfit2}. The WD
model spectrum matches fairly well to the observed continuum,
although the model flux peaks at around 1000\,\AA, bluewards of
the observed flux peak at 1060\,\AA. The best fit has a
temperature of 59,700K and \vsini\ of 240 $\VEL$, not very
different from that in the \fuse\ quiescent spectrum. In common
with the quiescent WD model fits, the model Lyman lines are
broader than the observed. However, the normalization at 163\,pc
gives \rwd=\EXPU{12.5}{8}{cm}, which would imply an unacceptably
low (0.35 \MSOL) WD mass.

We then fit model accretion disk spectra, as in
Section~\ref{ss:diskfits}.  We find that the standstill continuum
emission is reproduced well by an accretion disk model with
\mdot{disk}=\EXPU{6.9}{16}{\gs} at an inclination of 57\degr\ and
\mdot{disk}=\EXPU{1.1}{17}{\gs} at 68\degr\ (see Fig.\
\ref{f:standfit2}). The disk models are a good fit to the
continuum (the higher-inclination model is slightly better),
although they are unable to reproduce the absorption lines. The
disk model falls off steeply towards the Lyman limit, whereas the
data (and to a certain extent the WD models) level off before
dropping steeply to zero.

\section{Discussion}
\label{s:discus}

\subsection{The continuum emission in Quiescence, Standstill and Outburst}
\label{ss:discus1}

In quiescence, we find that the FUV continuum of Z Cam as observed
by \fuse\ is qualitatively well-described in terms of emission
from a hot metal-enriched WD atmosphere.  For log g=8.5, the best
fit single temperature models has a \twd\ of 57,200 K and \rwd\ of
\EXPU{4.4}{8}{(d/163 pc)\: cm}.  The mass of the WD in Z Cam is
estimated to be 0.99$\pm$0.15 \MSOL, which suggests \rwd\ of
\EXPU{5.8_{-1.8}^{+1.2}}{8}{cm}.  Thus the two measurements of
\rwd\ can be reconciled if the distance to Z Cam is
$214_{-44}^{+41}$ pc, where the errors here are calculated from
the uncertainty in the mass.  This is slightly larger than, but
wholly consistent with the astrometric distance of
$163_{-38}^{+68}$ pc.

If this interpretation is correct, then the WD in Z Cam is hot
when compared to the WDs in other CVs.  A recent compilation of
reliably determined WD temperatures in CVs by \cite{05araujo}
contains 34 objects (other than Z Cam) ranging in temperature from
9,500 K in the polar EF Eri to 50,000 K in the nova-like variable
and SW Sex star DW UMa.  A temperature of 57,200 K would make Z
Cam the hottest WD in a CV, excepting only V1500 Cyg, which is
still cooling from its nova outburst in 1975 and for which there
is a crude blackbody temperature estimate of 70,000 to 120,000 K
for the WD \citep{95schmidt}.

Is this reasonable?  The surface temperatures of WDs in CVs are
thought to be primarily determined by the accretion history of the
WD \citep{85sion}. This is because although the WDs in CVs emerge
from the common envelope phase as very hot WDs, the WDs cool to
temperatures of 4,500-6000 K in the 3-4 Gyr required to reach the
mass-exchanging stage.  Accretion reheats the surface layers of
the WD on the time scales of outbursts (weeks), as is evidenced
from observations of temperature changes in WDs in CVs such as U
Gem \citep{93long} and VW Hyi \citep{96gaensicke}. More
importantly,  accretion reheats the entire WD over longer time
scales (\POW{8}{years}), as the envelope and core adjust to the
increase in mass of the WD, an effect known as compressional
heating, and through nuclear burning of the accreted material
\citep{95sion,02townsley}. Recently, \cite{03townsley} have
determined the relationship between time-averaged accretion rate
and WD temperature.  For a temperature of 57,000 K, the required
rate is $\sim$\POW{18}{g\:s^{-1}} (with uncertainties of perhaps a
factor of two). (A somewhat lower value of \EXPU{4}{17}{g\:s^{-1}}
is obtained for 45,000 K, the temperature derived from the
combined \fuse\ + \iue\ spectrum.) Thus the accretion rate
required is high, although it may be consistent with the fact that
Z Cam is commonly in  (or near) the high state. Indeed, of the
four systems with WD temperatures greater than 40,000 K, MV Lyr,
TT Ari, RU Peg, and DW UMa, all but RU Peg are nova-like variables
and hence, like Z Cam, are systems where there are other
indications of high time-averaged accretion rates.  Given this, we
believe a temperature of 57,000 K is reasonable.

An alternative interpretation of the \fuse\ quiescent spectrum is
that the surface temperature of the WD in Z Cam is not uniform.
Our analysis shows that a somewhat better, but still not
statistically acceptable fit to the \fuse\ data can be obtained if
2T models are considered. Our best fit had a low temperature
component with $T_1$ of 26,300 K covering 71\% of the surface and
a high temperature component with $T_2$ of 71,700 K covering 29\%.
If the low temperature component represents energy loss from the
interior of the WD while the high temperature component is due to
recent or ongoing accretion, then 94\% of the luminosity of the WD
is due to recent accretion. Is this reasonable?

If the excess emission is due to ongoing accretion, then the
observed luminosity is $f\:GM_{\rm WD}\dot{M}/ \rwd $, where $f$
is the fraction of gravitational energy available that is
radiated. For the best 2T parameters, the excess luminosity is
\EXPU{1.9}{33}{erg~s^{-1}}, and the implied value of $\dot{M}$ is
\EXPU{8.3}{15}{g~s^{-1}/f}. This is a fairly large value for a
disk in quiescence, especially since $f\le0.5$, but is hard to
rule out because of our lack of good models for emission from
quiescent disks and the brightness of the WD itself.

On the other hand, the accretion belt hypothesis is that some of
the kinetic energy of material that reaches the boundary layer
between the disk and WD is stored in a rapidly rotating belt and
released slowly during the interoutburst period.  In standard
accretion disk theory, half of the gravitational energy of
material accreted by the WD is radiated by the disk and half
remains as kinetic energy when it enters the boundary layer.  For
Z Cam, with a disk accretion rate of \EXPU{3}{17}{g~s^{-1}}, the
disk luminosity is \EXPU{1.8}{34}{erg~s^{-1}}. For the outburst
that preceded the \fuse\ observation, Z Cam remained within 1
mag.\ of outburst maximum for about 4 days (see Fig.\
\ref{f:aavso}), and hence the total energy release was about
\EXPU{2.4}{40}{ergs}. The \fuse\ observation occurred about 11
days from the return to quiescence and, as noted previously, the
belt luminosity in the context of this hypothesis was still
\EXPU{1.9}{33}{erg~s^{-1}}. We do not have the coverage to
determine the time decay of the second component, but a
conservative lower limit is obtained by assuming constant
luminosity until the observation and zero from then on. By this
argument the energy stored in the rotating layer was at least
\EXPU{1.8}{39}{ergs}, or 8\% of the energy of the outburst. If one
assumes an exponential decay, then the energy release is $L(t)
e^{(t/\tau)} \tau$, where t is the time from outburst, L(t) the
luminosity at the time of our observation, and $\tau$ the decay
time constant. If $\tau$ was 11 days, then the total energy
released would be $2.7\times$ greater, or 20\% of the total
outburst energy.  In either case, the energy that must be stored
in the belt is large, though not energetically disallowed.

If the hot component is produced by short-term effects of
accretion, then the low T component reflects the long-term average
of accretion.  For a temperature of 26,300, this requires an
accretion rate of \EXPU{1.2}{17}{g~s^{-1}}.  This is lower than
the number inferred from 1T fits, but still substantial.

So how does one decide between the two models we have explored for
Z Cam in quiescence? The data are not definitive. The model
parameters that result from both models are substantially in
agreement with other known parameters of the system. \CHINU\ is
somewhat better for the 2T model than the uniform temperature
model. But \CHINU\ is much greater than 1 per degree of freedom in
the 2T model, showing that it is at best a step toward a correct
model of Z Cam in quiescence, and the systematic errors in the
\fuse\ spectrum are difficult to quantify.  The differences in the
two model spectra are quite small; there is no characteristic of
the spectrum that one can point to that requires a second
component.  Evidence that \vsini\ for the high temperature
component is high compared to the low T component is lacking.  In
U Gem \citep{93long,01froning}, single temperature models lead to
a physically implausible result, namely that the WD grows in
radius during quiescence. Since we do not have multiple
observations of Z Cam through a quiescent period, we do not know
whether single T models lead to a similar problem in Z Cam. There
are other well-observed systems, e.g. VW Hyi \citep{96gaensicke}
and WZ Sge \citep{04long}, which do not exhibit ``radius growth''
in the context of one temperature models, and so one cannot use
that as a guide. There is also no consensus theoretically.  While
\cite{94meyer} have argued that mass transfer from the disk is
higher immediately after outburst in order to explain UV delays in
DNe, the expected X-ray evidence of a gradual decline in hard
X-ray emission following outbursts is lacking. Furthermore the
X-ray data that exist indicates that the X-ray plasma that exists
in quiescence is not rotating rapidly \citep{02szkodyb,04mauche},
and therefore it is hard to understand how it could produce a
heated annular region around the surface. While \cite{78kip}
postulated the existence of accretion belts and described an
instability that might release energy in an accretion belt, there
are to our knowledge no detailed calculations of expected
properties of the emitting region and no attempt to calculate the
luminosity of the rotating component with time. Given these
considerations, our answer to the question of which model is most
likely correct is to appeal to Occam's Razor and assert that the
simpler 1T model provide a good physical description of the data
and is therefore is the best-bet physical model to pursue.

In outburst, \cite{97knigge} showed that the shape of the FUV
(900--1800\AA) continuum of Z~Cam could be modeled as an accretion
disk with $\mdot{disk}\simeq\EXPU{3}{17}{\gs}$ at an inclination
of 57\degr\ and distance of 170 pc (very close to the best
astrometric distance of 163 pc). In the high state, the fits are
self-consistent.  In standstill we find
$\mdot{disk}\sim\EXPU{7_{-3}^{+4}}{16}{\gs}$. Approximately, then,
there appears to be at least an order of magnitude increase in
\mdot{} from quiescence to standstill, while \mdot{outburst} is an
order of magnitude higher still. Our derived \mdot{standstill}
gives an upper limit on \mdot{crit}, as it is a reasonable
assumption that \mdot{standstill} is also the rate of mass flow
from the secondary at that time, which must be greater than
\mdot{crit}. \citet{01buat} calculate
$\mdot{crit}=3\times10^{17}$\gs, using the formula of
\citet{98hameury} and taking into account heating by the
stream--disk impact (or \EXPU{2}{17}{\gs} if tidal dissipation is
also included). Our upper limit of $\mdot{crit}\la
\EXPU{2}{17}{\gs}$ is just consistent with this and with the upper
limit of $\mdot{crit}\la \EXPU{3.2}{17}{\gs}$ derived by
\citet{00baraffe}.

The comparison of the BEFS standstill spectrum and the \fuse\/
spectrum of Z~Cam in quiescence illustrates the problem of
separating a WD spectrum from that of a disk. Except for the flux
level the qualitative differences between the BEFS spectrum and
the \fuse\ spectrum are relatively small. In both states, the
continuum slope is almost the same and, other than a slight
increase in ionization temperature, their line spectra are very
similar. Yet, we find that in quiescence a WD alone can make up
the continuum flux, whereas in standstill an accretion disk is the
best model for the continuum flux. Ultimately, the key model
parameter that has enabled us to fit one model and reject others
is the normalization required to match the observed and the
modeled flux levels, which in turn depends on the WD mass and the
distance from the sun. Without a reasonably accurate knowledge of
these parameters it would be impossible to judge between a wide
range of models that reproduce the continuum slope and line
spectrum equally well.
\subsection{Phase-dependent variability of the spectral lines}
In the time-resolved spectra of Z~Cam in quiescence, we have seen
enhanced line absorption between phase 0.65 and 0.81, which covers
the full range of atomic species and excitation levels. The
enhanced absorption is very well centered on the mean line
profile, which, in turn, is centered 100--200\kms\ to the red of
the transition rest velocity. The change in absorption width is
greatest in those lines that are weak or absent from the mean
spectrum, but also encompasses lines which are thought to be
optically thick in the mean spectrum, leaving only the N\IVl924,
O\VIl$\lambda\,$1032,1038 and C\IIIl977 and 1176 lines unchanged.

Flux variations at around phase 0.5--0.8 are seen in a selection
of disk-accreting systems, in phenomena such as dips in the UV
\citep{97mason} and X-ray \citep{93hellier} light curve of
low-mass X-ray binaries and CV, and humps in the optical light
curves of CV \citep[e.g.][]{89wood}.  This is typically attributed
the interaction of the accretion stream from the secondary with
the outer edge of the accretion disk. This can result in a
luminous `bright spot' at the outer edge of the disk, which is
thought to be responsible for the optical light curve modulation.
Also, the shock of the in-falling stream can cause the disk to
bulge outwards at smaller radii or cause the stream to be
deflected upward and arc over the surface of the accretion disk
\citep[e.g.][]{98armitage}. This can cause phase dependent
absorption as colder material in the accretion stream moves in
front of the continuum source.

The additional absorption seen in Z~Cam is qualitatively compatible
with some sort of stream-disk interaction.  The enhanced absorption is
narrow and centered on the mean line profile, indicative of matter
that is moving across, rather than parallel to the line of
sight. There is no corresponding change in the ionization state and
the increase in absorption affects those transitions that appear to be
optically thick in the mean spectrum. This is consistent with a
disk bulge or stream temporarily moving to occult a larger area of the
continuum source, or to place a greater column density of absorbing
ions into the line of sight. Also, the 0.65--0.81 phasing is in the
phase range that stream-overflow effects are thought to be seen. In
these respects Z~Cam compares well with the
\fuse\/ spectrum of U~Gem in outburst \citep{01froning}, which
was observed to show strong enhanced absorption at phase 0.53 to
0.79. The line variability in U~Gem is of a similar strength to that
seen in Z~Cam, i.e. residual line core flux reducing to as little as
40\,\% of its mean value in some lines, with the O\VI\ lines being the
least affected, although the N\IVl924, C\IIIl1176 and other weak lines
participate in the variation in U~Gem.

Smooth particle hydrodynamical models of the disk-stream
interaction are now being produced
\citep[e.g.][]{96armitage,98armitage,01kunze}. These models make
predictions of the observational qualities which we are now able
to set against the evidence for stream-related variability seen in
the spectrum of Z~Cam. The models are generally in agreement that
stream overflow can occur at low mass-accretion rates, i.e.
quiescent disks. \citet{98armitage} predict that efficient cooling
of the hot-spot region can occur in systems with low
mass-accretion rates ($\mdot{}\la10^{-9}\MSOL \: yr^{-1}$, or $\la
\POW{16}{\gs}$), allowing material to overflow the inner disk in a
coherent stream, whereas, in higher-\mdot\ systems the stream
impact region tends to push the disk into a bulging shape. Using a
different method, \citet{99hessman} concurs with Armitage \& Livio
that up to $\sim10\,$\% of the matter in the accretion stream from
the secondary is able to overflow the disk.

If the phase-dependent absorption is a result of stream-overflow,
it may have implications for the estimated inclination of the
system. Logically, for absorption by the stream, the absorption
strength should have a strong correlation with inclination angle.
At higher inclinations, material above the disk plane presents a
higher column density to the bright continuum regions, i.e. the WD
and inner disk. \citet{01kunze} estimate that in outburst,
enhanced absorption can be seen in systems of inclinations down to
65\degr, based on the maximum scale height above the disk attained
by material in the accretion stream. However, they predict that,
at low \mdot{}, material is not as strongly deflected upward as
for high \mdot\ and that an inclination angle of at least 75\degr\
is needed before the overflowing stream can cause accretion dips.
At an inclination of only $57\pm11$\degr \citep[as taken
from][]{83shafter}, Z~Cam would fall below this lower limit.
Furthermore, an inclination above the current upper limit of
68\degr\ for Z Cam would produce observable eclipses and therefore
can be ruled out. However, given the similarity of the
time-dependent absorption to that seen in U~Gem ($i=67\pm3$\degr),
an inclination angle near 68\degr seems far more plausible that
the opposite extreme of 48\degr.


\section{Conclusions}
We have obtained observations with the \fuse\/ satellite of the
dwarf nova, Z~Cam, during a period of quiescence. The spectrum is
characterized by a fairly flat continuum, with a peak in the flux
at 1000\AA. The line spectrum covers a broad range of atomic
species and ionization stages and is dominated by absorption lines
of FWHM $<1000$\kmsd. Qualitatively, the spectrum can be
self-consistently described in terms of emission from the WD with
a temperature of 57,000 K. Modest improvements in the fits are
obtained with 2T models in which most of the emission comes from a
small fraction (29\%) of the surface heated to $71,700\,$K. The
remainder is at 26,300\,K. We favor the uniform temperature model
for Z Cam, largely because the data do not require anything more
complicated. There are emission lines (e.g. C\IIIl1176) that do
not come from the WD. The widths of these lines are consistent
with a disk origin.

In the phase-resolved quiescent spectrum we have observed
transient enhanced line absorption. From the 0.65--0.81 phasing of
this absorption and its effect on a wide range of ionization
species, we attribute it to material raised from the disk plane
into the LOS, due to interaction of the accretion stream with the
disk, or the stream itself moving over the disk. Such effects may
be easier to understand if inclination of Z Cam is close to the
upper limit of 68\degr.

The standstill continuum is best described by an optically-thick
disk accreting at a rate of \EXPU{7_{-2}^{+4}}{16}{\gs}, depending
on the actual inclination of Z Cam. The continuum slope and the
line absorption spectra differ little between quiescence and
standstill, despite that, during standstill, an accretion disk
creates (most of) the continuum flux. The qualitative similarity
of the two spectra highlights the difficulty of separating WD from
disk signatures without detailed model fitting and, also,
underlines the reliance of the model fits on an secure knowledge
of the distance and the WD mass.

The high temperature of the WD in Z Cam and the estimates of mass
accretion rate in standstill and outburst are all consistent with
a mass transfer rate from the secondary star that is higher than
in normal DNe. Additional FUV observations of Z Cam in the period
following a normal outburst are highly desirable both to better
understand the physics of  standstill systems and to cleanly
distinguish between single and multi-temperature models of the WD
in Z Cam.



\singlespace \acknowledgements In this research, we have used, and
acknowledge with thanks, data from the AAVSO International Database,
based on observations submitted to the AAVSO by variable star
observers worldwide. We also gratefully acknowledge the financial
support from NASA through grant NAG 5-11885.  All of the data
presented in this paper was obtained from the Multimission Archive at
the Space Telescope Science Institute (MAST). STScI is operated by the
Association of Universities for Research in Astronomy, Inc., under
NASA contract NAS5-26555. Support for MAST for non-HST data is
provided by the NASA Office of Space Science via grant NAG5-7584 and
by other grants and contracts. We thank the anonymous referee for
their helpful comments and suggestions.



\pagebreak


\figcaption{V-magnitude light curve of Z Cam from 2001 December 16
to 2002 March 6 (AAVSO, Mattei J. A., 2002,
  Observations from the AAVSO international database, private
  communication). The date of our FUSE observation (February 9 2002,
  MJD 52314.25) is marked with a vertical line. \label{f:aavso}}

\figcaption{Mean \fuse\/ spectrum of Z~Cam created from the
exposure time-weighted mean of all 43
exposures.\label{f:spectrum}}

\figcaption{DA WD model fit to the night-only mean quiescent
spectrum of Z~Cam. The upper panel plots the observed spectrum and
best-fit model (in red); the lower panel plots the residual
spectrum after the model has been subtracted and 1$\sigma$ errors
in the observed spectra (in blue). Only metal-line free portions
of the spectrum were included in the DA model fits, the excluded
regions plotted in grey.  Model parameters are given in
Table~\ref{t:fits}.}\label{f:dafit}

\figcaption{Solar abundance model WD fits to the mean \fuse\
spectrum of Z~Cam. In the upper panel the solid (red) line plots
the best statistical-fit single temperature model and the dashed
(blue) line plots our  two-temperature model. The lower panel
plots the difference spectrum of the observed and
single-temperature model flux, with (1-$\sigma$) errors in blue.
The grey portions of the plotted spectrum were not considered in
fitting the data.  Model parameters are given in
Table~\ref{t:fits}.\label{f:wdfit1}}

\figcaption{Solar abundance WD model fits to the merged \fuse\ and
\iue\ spectra of Z~Cam. In the upper panel the solid line plots
the best statistical-fit single temperature model and the dashed
(blue) line plots our two-temperature model. The lower panel plots
the difference spectrum of the observed and single-temperature
model flux, with errors in blue. The grey portions of the spectrum
that have been masked out from the fits. Model parameters are
given in Table~\ref{t:fits}.\label{f:dafuseiue}}

\figcaption{ The best-fit accretion disk models compared to  the
quiescent spectrum of Z Cam. The solid (red) line indicates the
57\degr\ model and the long-dashed (blue) line indicates the
68\degr\ model with the distance fixed at 163 pc. The dashed
(black) line is the best fit when the normalization (or
equivalently the distance) is a free parameter. Model parameters
are given in Table~\ref{t:fits} \label{f:diskfit1}}

\figcaption{Disk fits to the merged \fuse\ and \iue\ spectra of
Z-Cam. The solid (red) curve is the best fit when the distance is
constrained to be 163 pc and the inclination is 57\degr. The
long-dashed (blue) curve is similar but for an inclination of
68\degr.  The dashed (black) curve is the best fit when the
normalization (or equivalently the distance) is a free parameter.
As previously, the lower panel is difference between the best
57\degr\ model and the data. \label{f:fuse_iue_disk}}

\figcaption{Line profile fits to the
S\IV$\,\lambda\lambda\,$1063,1072, Si\IIIl1108--1113 and
C\IIIl1175 lines, after the continuum has been normalized. The
lines are fit using Gaussian emission and absorption line profiles
using the {\sc dipso} software. When the multiple components of a
line, e.g. S\IVl1063 and S\IV1072 can be separated, one Gaussian
absorption and one Gaussian emission profile has been used for
each component. \label{f:emission}}

\figcaption{A phase-resolved segment of Z~Cam FUV spectrum showing
in the range 1005--1045\AA. The main features are due C\IIIl1010,
S\IIIl1012--1021, \lyb\ and O\VIl$\lambda\,$1032,1037 lines. The
phased spectra are created by summing those exposures which, all
or mostly, fall inside the given phase range. The grey line is the
time-averaged spectrum from all 43 radial-velocity corrected
exposures.\label{f:variation1}}

\figcaption{Selected regions of the Z~Cam spectra, showing the
mean spectrum between phases 0.65 and 0.81, over-plotted on the
mean spectrum from the full observation (gray line). The spectra
are summed from the individual exposures, after they have been
radial velocity corrected according to the ephemeris of
\citet{95thorstensen}.\label{f:addabs}}

\figcaption{Time-averaged spectrum of Z~Cam in standstill as
observed by the BEFS on-board the \orfeus\ satellite, plotted with
the quiescent \fuse\ spectrum (from those exposures recorded
mostly during spacecraft night). The flux scale is
$10^{-13}$\funits\ for the \fuse\ spectrum and $10^{-12}$\funits\
  for the BEFS spectrum, the latter is also offset upward by
  \EXPU{2}{-12}{\funits}, so that the standstill spectrum lies above
the quiescent spectrum. In the lower panel is plotted the ratio of
the standstill flux to the quiescent flux.\label{f:standstill}}

\figcaption{Solar abundance WD model fits to the BEFS spectrum of
Z~Cam in standstill. In the upper panel, the solid (red) line
plots the best-fit model. Model parameters are given in
Table~\ref{t:fits}.\label{f:standfit2}}

\figcaption{Disk model fits to the BEFS spectrum of Z~Cam in
standstill. In the upper panel, the solid (red) line plots the
best fit accretion disk model spectrum at 57\degr\ and the
long-dashed (blue) lines plots our best disk model at 68\degr. The
dashed (black) line is the best fit when the normalization is not
constrained. Model parameters are given in
Table~\ref{t:fits}.\label{f:standfit}}


\begin{deluxetable}{lcc}
\tablewidth{0pt}
\tablecaption{Adopted system parameters for Z~Cam.\label{t:param}}
\tablehead{
\colhead{Parameter} &
\colhead{Value} &
\colhead{Reference}
}
\startdata
$P_{\rm orb}$ (d)   & $0.2898406(2)$    &   1   \\
$K_1$ (\kmsd)   & $135\pm9$ &   1   \\
$i$ (\degr) & $57\pm11$ &   2   \\
$M_1$ (\MSOL)   & $0.99\pm0.15$ &   2   \\
$M_2/M_1$   & $0.71\pm0.10$ &   2   \\
$d$ (pc)        & $163^{+68}_{-38}$&    3   \\
\enddata
\tablerefs{(1) \citealt{95thorstensen} (2) \citealt{83shafter} (3)
\citealt{03thorstensen}}
\end{deluxetable}

\begin{deluxetable}{lcccc}
\tablewidth{0pt} \tablecaption{Spectral lines identified in the
\fuse\/ spectrum of Z~Cam in quiescence. \label{t:lines}}
\tablehead{ \colhead{Ion} & \colhead{$\lambda_{\rm lab}$ } &
\colhead{EW)} & \colhead{FWHM }
\\
\colhead{~} & \colhead{(\AA)} & \colhead{(\AA)} & \colhead{
(\kms)} } \startdata
N\IV* + H\I...   & 922--924; 923.2  & 2.02$\pm$0.10     & 1030$\pm$65   \\
H\I...              & 930.748       & 0.6$\pm$0.1   & 570$\pm$75     \\
S\VI...            & 933.4      & 0.79$\pm$0.06     & 565$\pm$30    \\
H\I...            & 937.8       & 1.1$\pm$0.1   & 545$\pm$45    \\
S\VI...            & 944.5      & 0.82$\pm$0.06     & 470$\pm$20 \\
H\I + P\IV...    & 949.7; 950.7     & 1.80$\pm$0.1  &900$\pm$60\\
H\I...              & 972.5         & 1.09$\pm0.08$     & 600$\pm$50\\
C\III...             & 977.0        & 1.14$\pm$0.06     & 590$\pm$40\\
He\II* + N\III...& 989.8; 992.3         & 2.88$\pm$0.09     & 1150$\pm$40\\
Si\III*...           & 997.4        &$0.17\pm0.02$  & $300\pm20$\\
C\II*...            & 1010.0        & 0.20$\pm$0.04     & 420$\pm$90\\
S\III...             & 1012.5       & 0.09$\pm$0.02     & 240$\pm$60\\
S\III...             & 1015.5       & 0.40$\pm$0.02     & 410$\pm$30\\
S\III...             & 1021.1       & 0.56$\pm$0.03     & 500$\pm$30 \\
\lyb...             & 1025.7        & 1.0$\pm$0.2   & 470$\pm$90\\
O\VI...            & 1031.9         & 1.57$\pm$0.03     & 610$\pm$20\\
O\VI...            & 1037.6         & 1.88$\pm$0.03     & 700$\pm$20\\
S\IV...            & 1062.7         & 0.75$\pm$0.07     & 480$\pm$30\\
S\IV...            & 1073.0         & 0.73$\pm$0.05     & 480$\pm$30\\
S\III*...            & 1077.2       & 0.11$\pm$0.02     & 280$\pm$61\\
He\II* + N\II...  & 1084.9; 1085.0  & 1.9$\pm$0.2   & 1020$\pm$80 \\
Si\III*...           & 1108.4; 1110.0   & 1.65$\pm$0.03 & 890$\pm$20 \\
Si\III*...           &1113.2        & 0.98$\pm$0.02 & 500$\pm$20 \\
P\V...             & 1118.0         & $<0.1$        & $<500$ \\
Si\IV*...          & 1122.5         & 0.6$\pm$0.1   & 730$\pm$50\\
P\V + Si\IV*... & 1128.0; 1128.3    & 0.8$\pm$0.1   & 850$\pm$80 \\
C\III*...            &1175.3        & 3.28$\pm$0.06     & 940$\pm$30\\
\enddata
\end{deluxetable}

\begin{deluxetable}{cccccccc}
\tablecaption{Model Fits to Z Cam in quiescence and standstill
\label{t:fits}} \tablehead{\colhead{Model~type} &
 \colhead{$\rwd$} &
 \colhead{$\twd$} &
 \colhead{v~sin(i)} &
 \colhead{$\mdot{disk}$} &
 \colhead{D$^a$} &
 \colhead{\CHINU} &
 \colhead{N}
\\
\colhead{~} &
 \colhead{(10$^{8}$~cm)} &
 \colhead{(K)} &
 \colhead{($\VEL$)} &
 \colhead{(10$^{15}$~g~s$^{-1}$)} &
 \colhead{(pc)} &
 \colhead{~} &
 \colhead{~}
} \scriptsize \tablewidth{0pt}\startdata
\multicolumn{8}{c}{\sc QUIESCENCE~(FUSE)}\\
DA &  4.3 &  58,100 &  - &  - & 163 &  ~6.7 &  ~672   \\
WD &  4.4 &  57,200 &  330 &  - &  163 &  10.1 &  1146 \\
2WD &  6.0~(71\%/29\%) &  26,300/71,700 &  170/370 &  - &  163 &  ~9.4 &  1146 \\
Disk~(46\degr) &  - &  - &  - &  ~~6.6 &  163 &  32.1 &  1146 \\
Disk~(57\degr) &  - &  - &  - &  ~~9.1 &  163 &  27.1 &  1146 \\
Disk~(68\degr) &  - &  - &  - &  ~12.9 &  163 &  22.0 &  1146 \\
Disk~(57\degr) &  - &  - &  - &  180.0 &  710 &  11.9 &  1146 \\
Disk~(46\degr) \& WD &4.0    &62,000 &306    & ~~1.2 & 163 &  9.8    &1146\\
Disk~(57\degr) \& WD &4.0    &62,000 &295    & ~~1.7 & 163 &  9.8    &1146\\
Disk~(68\degr) \& WD &3.9    &62,000 &293    & ~~2.5 & 163 &  9.8    &1146\\
\multicolumn{8}{c}{\sc QUIESCENCE~(FUSE~\&~IUE)}\\
WD &  6.0 &  44,600 &  - &  - &   163 &  ~8.2 & ~471 \\
2WD &  18.4(97\%/3\%) &  17,600/82,400 &  - &  - &  163 &  ~4.9 &  ~471 \\
Disk~(46\degr) &  - &  - &  - &  ~~6.0 &  163 &  11.1 &  ~471 \\
Disk~(57\degr) &  - &  - &  - &  ~~8.7 &  163 &  ~9.6 &  ~471 \\
Disk~(68\degr) &  - &  - &  - &  ~12.9 &  163 &  ~8.1 &  ~471 \\
Disk~(57\degr) &  - &  - &  - &  182.0 &  710 &  ~5.8 &  ~471 \\
Disk~(46\degr) \& WD &3.0    &82,000     & -  & ~~3.0 & 163 &4.7&~471\\
Disk~(57\degr) \& WD &3.0    &81,000     & -  & ~~4.0 & 163 &5.0&~471\\
Disk~(68\degr) \& WD &3.0    &81,000     & -  & ~~6.3 & 163 &5.4&~471\\
\multicolumn{8}{c}{\sc STANDSTILL~(ORFEUS)}\\
WD &  12.5 &  59,700 &  240 &  - &  163 &  ~2.6 &  1505 \\
Disk~(46\degr) &  - &  - &  - &  ~43.0 &  163 &  ~2.7 &  1505 \\
Disk~(57\degr) &  - &  - &  - &  ~69.0 &  163 &  ~2.6 &  1505 \\
Disk~(68\degr) &  - &  - &  - &  112.0 &  163 &  ~2.5 &  1505 \\
Disk~(57\degr) &  - &  - &  - &  195.0 &  270 &  ~2.4 &  1505 \\
\tablenotetext{~^a}{ A value of 163 pc implies the distance was
fixed at this value.}
\enddata
\label{fits}
\end{deluxetable}

\begin{deluxetable}{lcccc}
\tablewidth{0pt} \tablecaption{Spectral lines identified in the
BEFS spectrum of Z~Cam in standstill.\label{t:standstill}}
\tablehead{ \colhead{Ion} & \colhead{$\lambda_{\rm lab}$} &
\colhead{EW } & \colhead{FWHM }
\\
\colhead{~} & \colhead{(\AA)} & \colhead{(\AA)} & \colhead{
(\kms)
}

} \startdata
H\I...                  & 930.748   & $0.68\pm0.09$ & $220\pm30$ \\
S\VI...             & 933.4         & $1.20\pm0.19$ & $460\pm100$ \\
H\I...                  & 937.8         & $0.81\pm0.13$ & $590\pm120$\\
S\VI...             & 944.5         & $1.35\pm0.11$ & $750\pm130$\\
H\I\ + P\VI...      & 949.7         & $1.68\pm0.12$ &$730\pm60$\\
H\I...                  & 972.5         & $1.12\pm0.07$ &$490\pm40$\\
C\III...                & 977.0         & $1.84\pm0.08$ &$760\pm40$ \\
N\III\ + Si\III*... & 989.8; 993.5  & $2.13\pm0.10$ &$1380\pm100$\\
Si\III*...              & 994.8         & $<0.2$        &$<300$ \\
Si\III*...              & 997.4         & $0.24\pm0.03$     &$310\pm50$ \\
S\III...                & 1012.5        & $<0.1$        & $<300$\\
S\III...                & 1015.5        & $<0.1$        & $<300$\\
S\III...                & 1021.1        & $0.22\pm0.04$ &$320\pm70$\\
\lyb...                 & 1025.7        & $0.76\pm0.07$ & $320\pm40$\\
O\VI...             & 1031.9        & $2.07\pm0.09$ & $1000\pm50$\\
O\VI...             & 1037.6        & $2.40\pm0.10$ & $745\pm40$\\
S\IV...             & 1062.7        & $0.98\pm0.45$ & $700\pm40$\\
S\IV...             & 1073.0        & $0.94\pm0.45$ & $460\pm30$\\
He\II* + N\II...    & 1084.9; 1085.0    & $0.58\pm0.08$ & $1110\pm150$\\
Si\III*...              & 1108.4; 1110.0    & $0.50\pm0.04$ & $420\pm40$\\
Si\III*...              &1113.2         & $0.56\pm0.06$ & $380\pm60$\\
P\V...              & 1118.0        & $0.92\pm0.05$ & $580\pm40$ \\
Si\IV*...           & 1122.5        & $0.53\pm0.07$ & $660\pm90$\\
P\V\ + Si\IV*...    & 1128.0; 1128.3    & $0.76\pm0.07$     & $740\pm70$\\
\enddata
\end{deluxetable}

\begin{figure}
\plotone{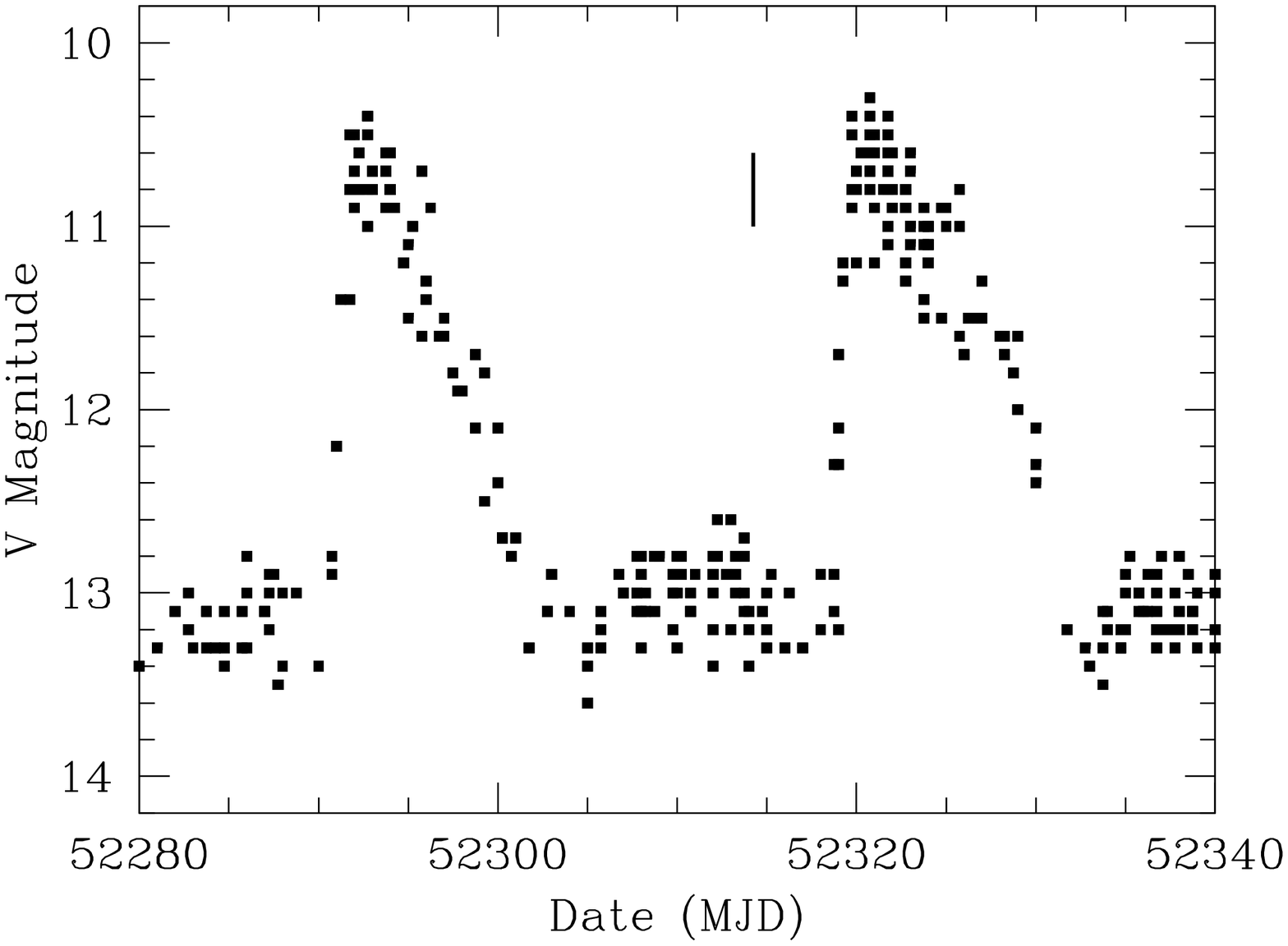}
\end{figure}

\begin{figure}
\plotone{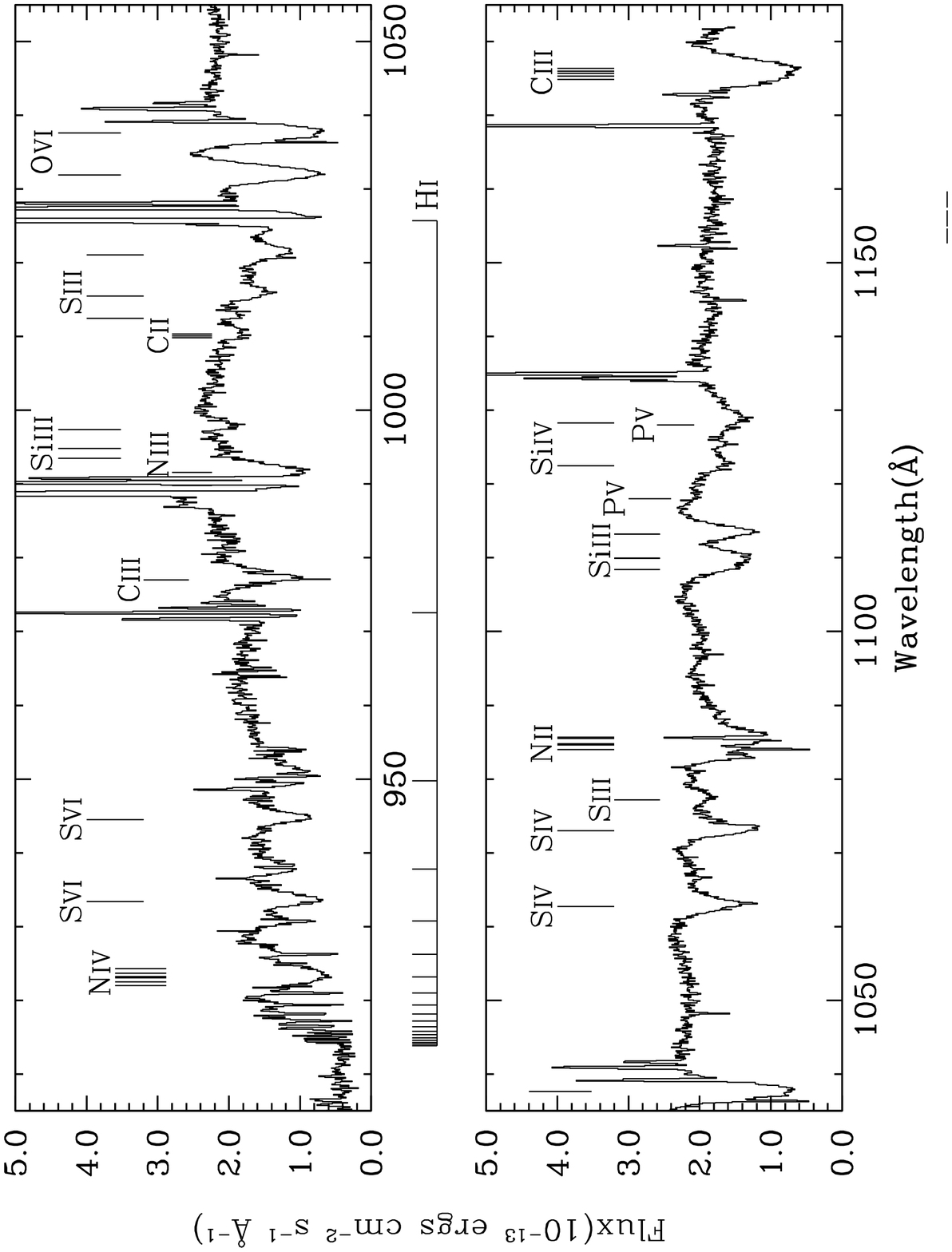}
\end{figure}

\begin{figure}
\plotone{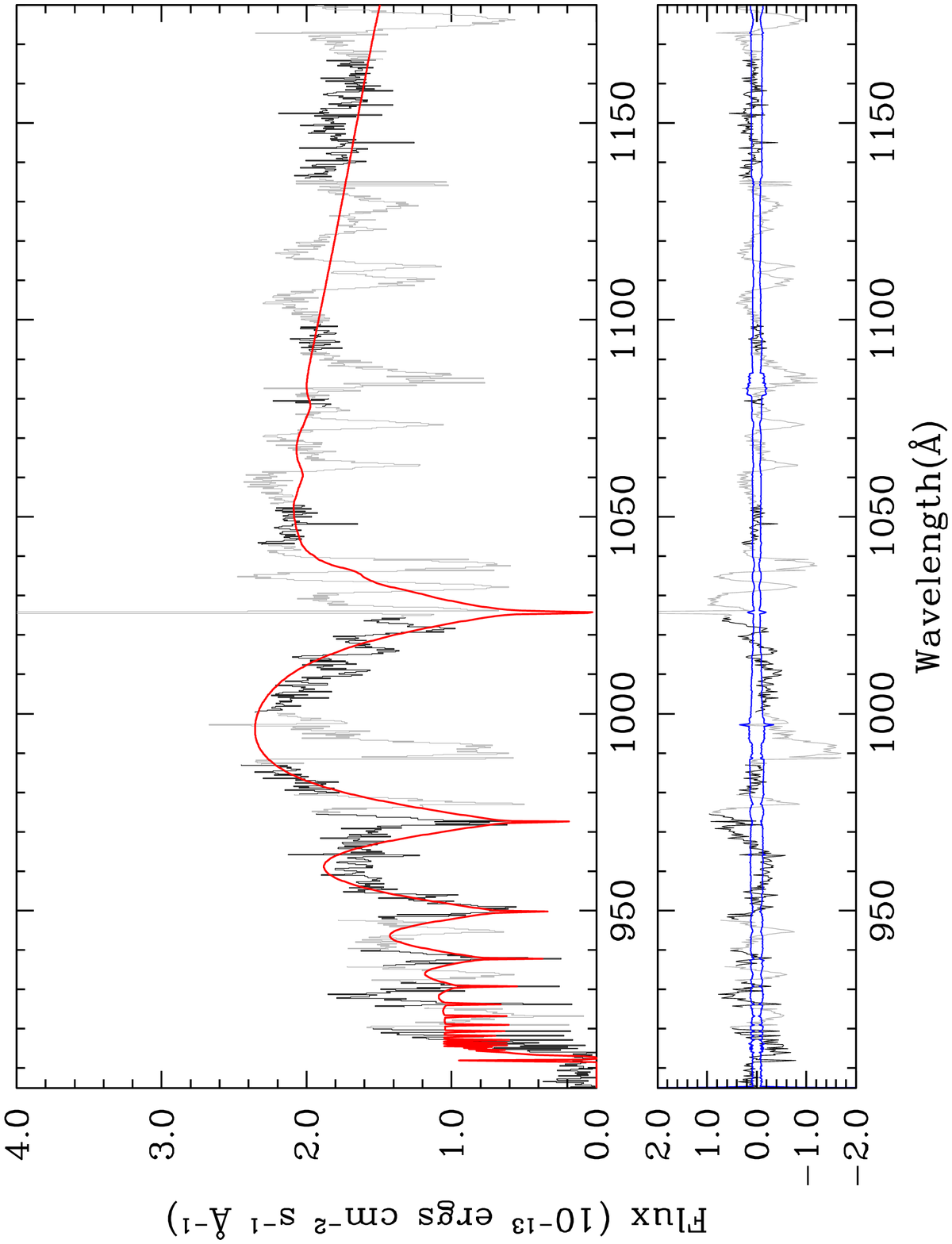}
\end{figure}

\begin{figure}
\plotone{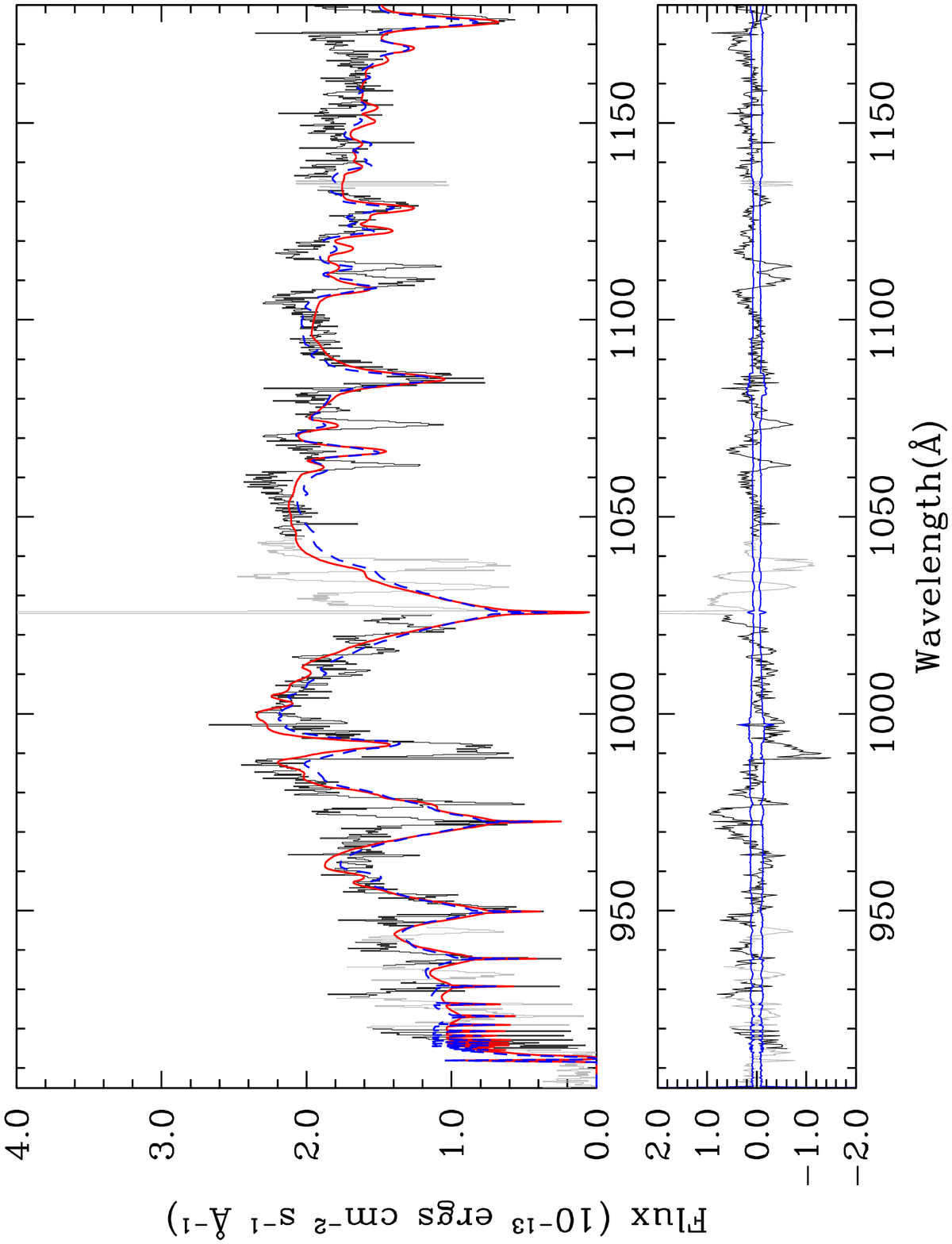}
\end{figure}

\begin{figure}
\plotone{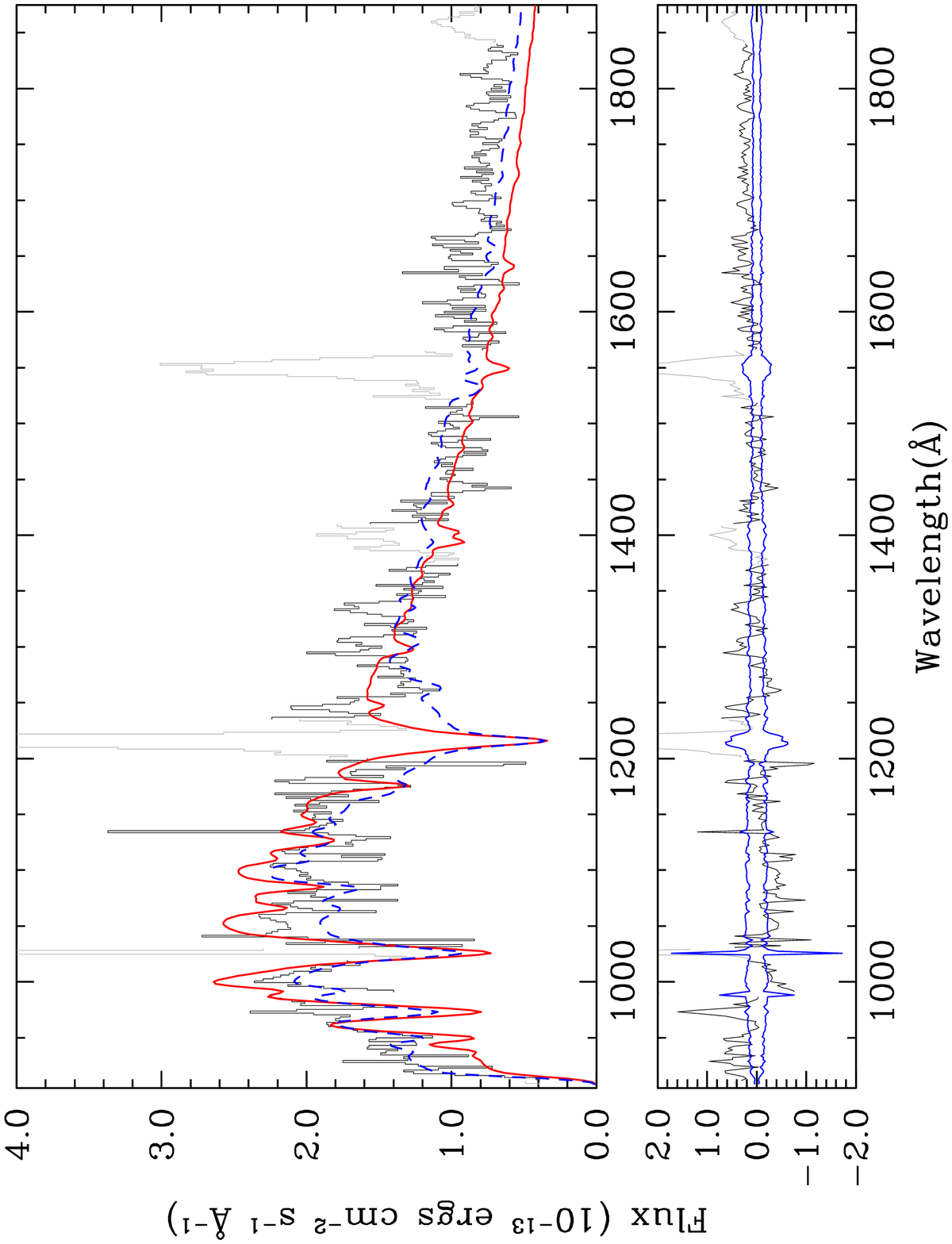}
\end{figure}

\begin{figure}
\plotone{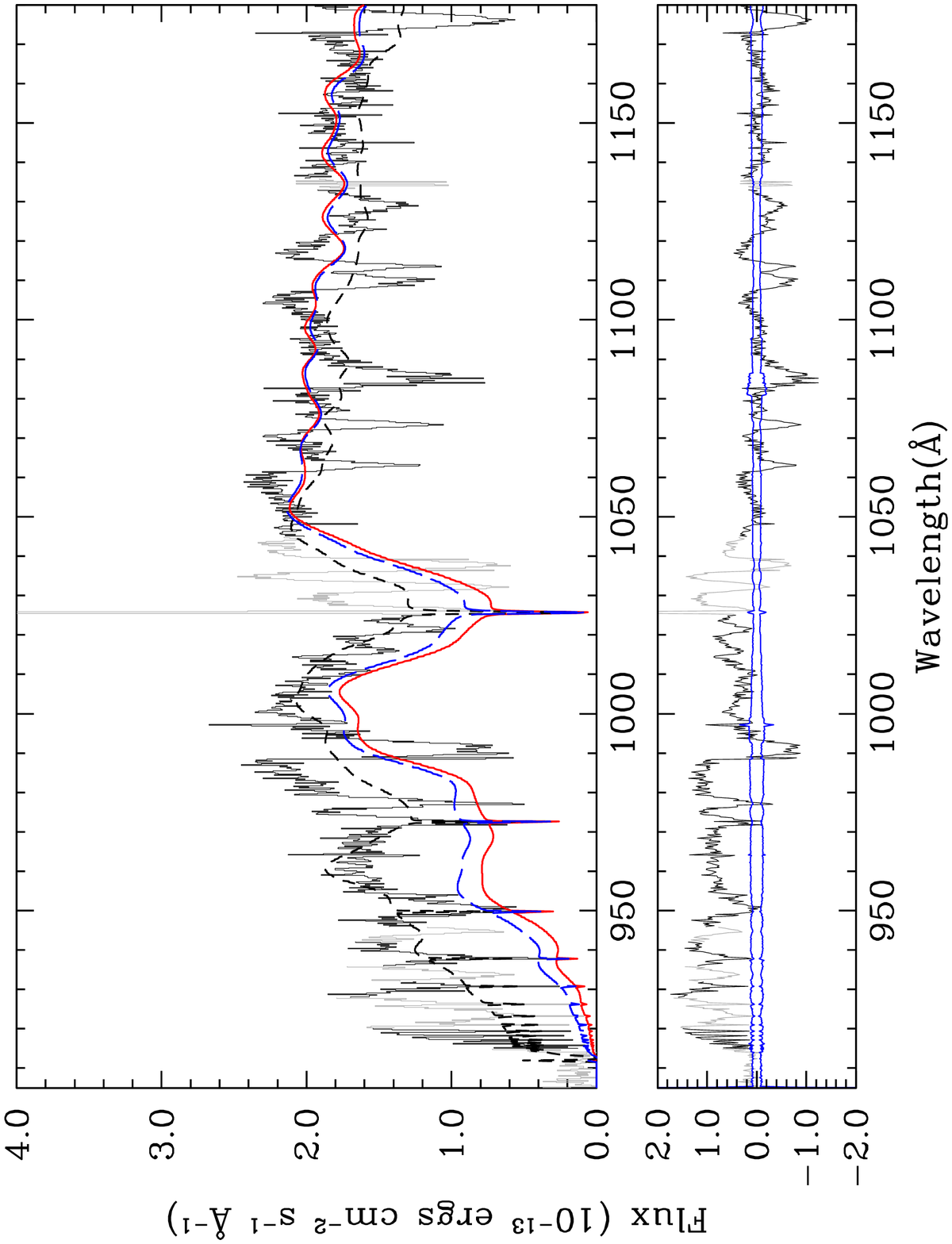}
\end{figure}

\begin{figure}
\plotone{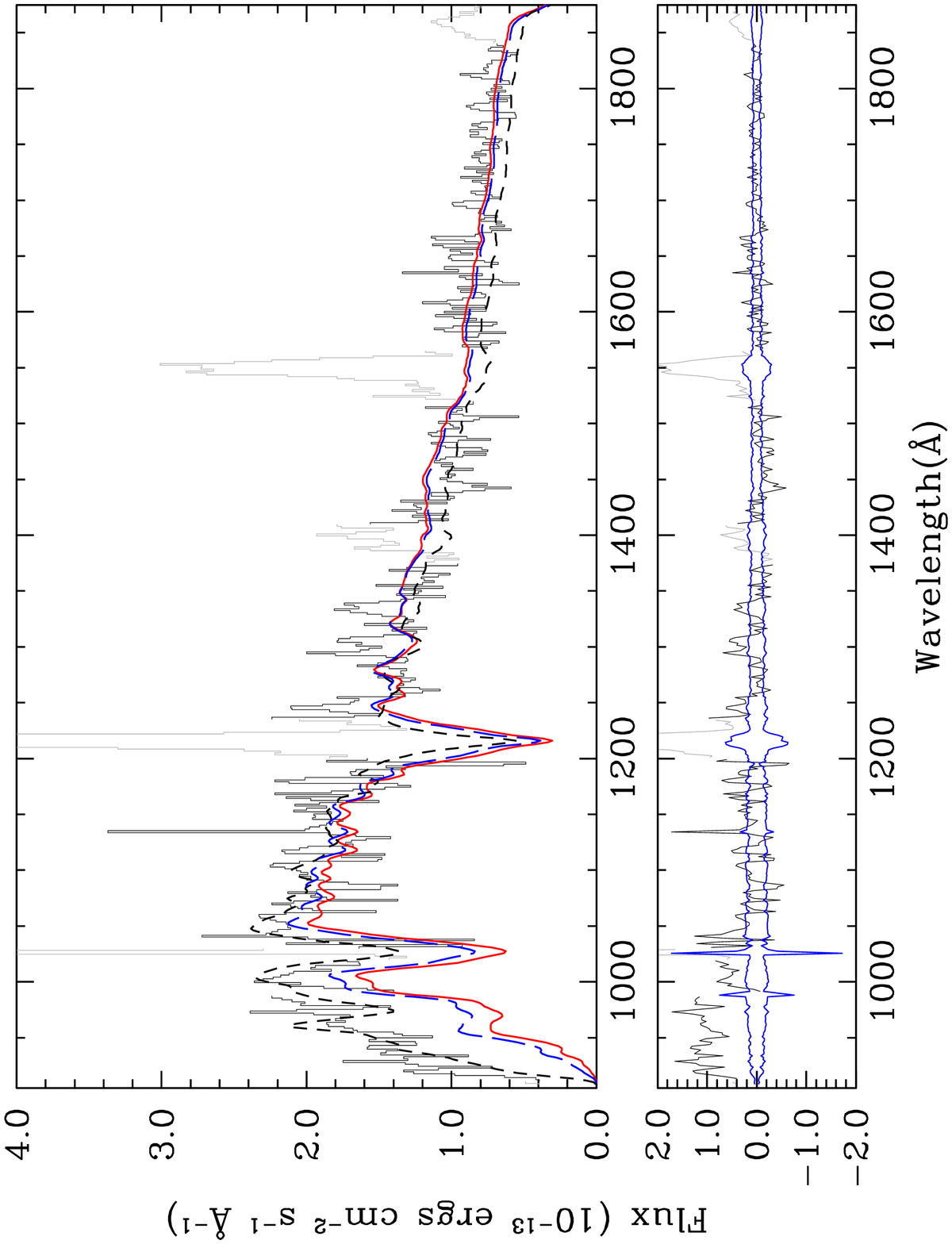}
\end{figure}

\begin{figure}
\plotone{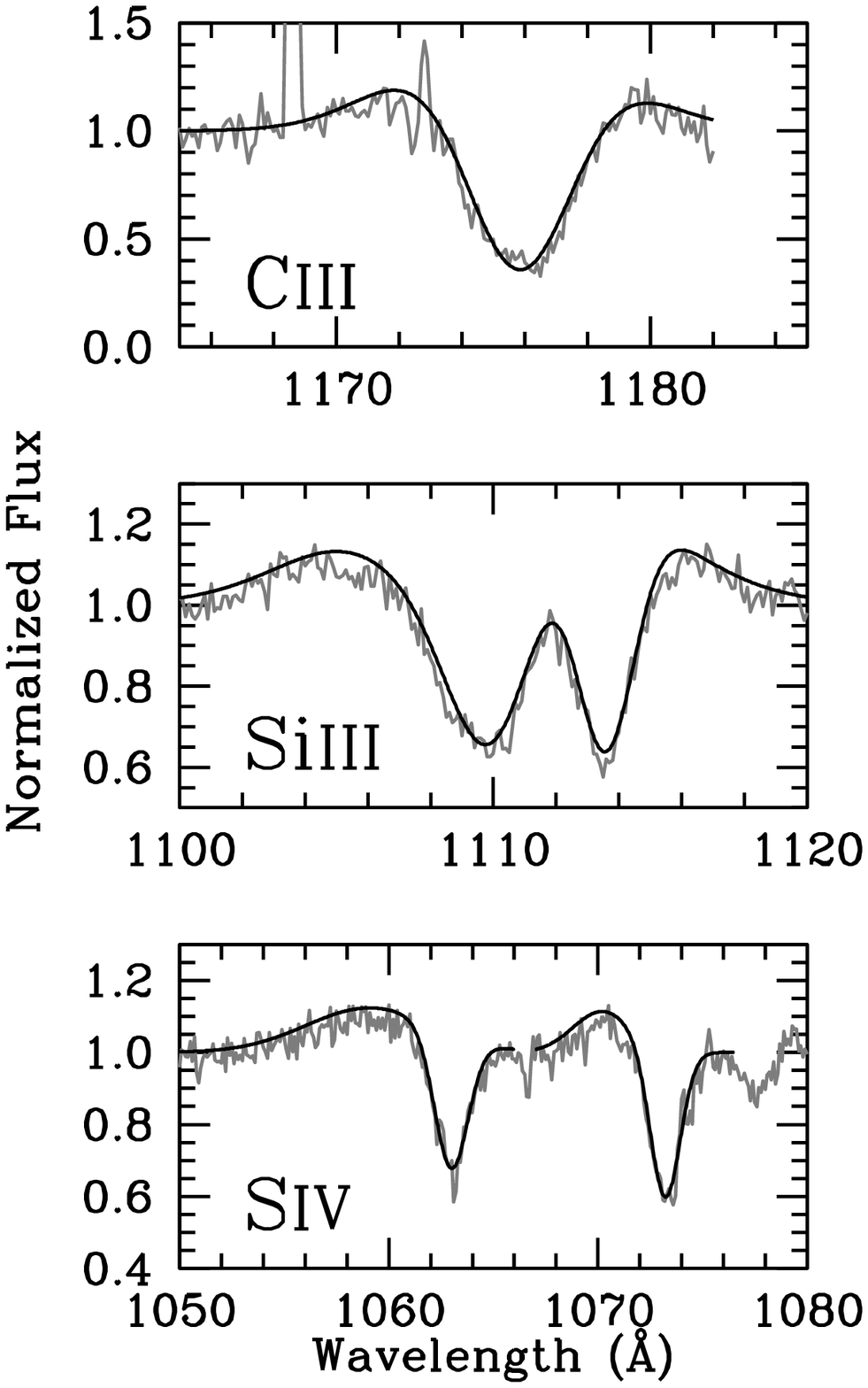}

\end{figure}

\begin{figure}
\plotone{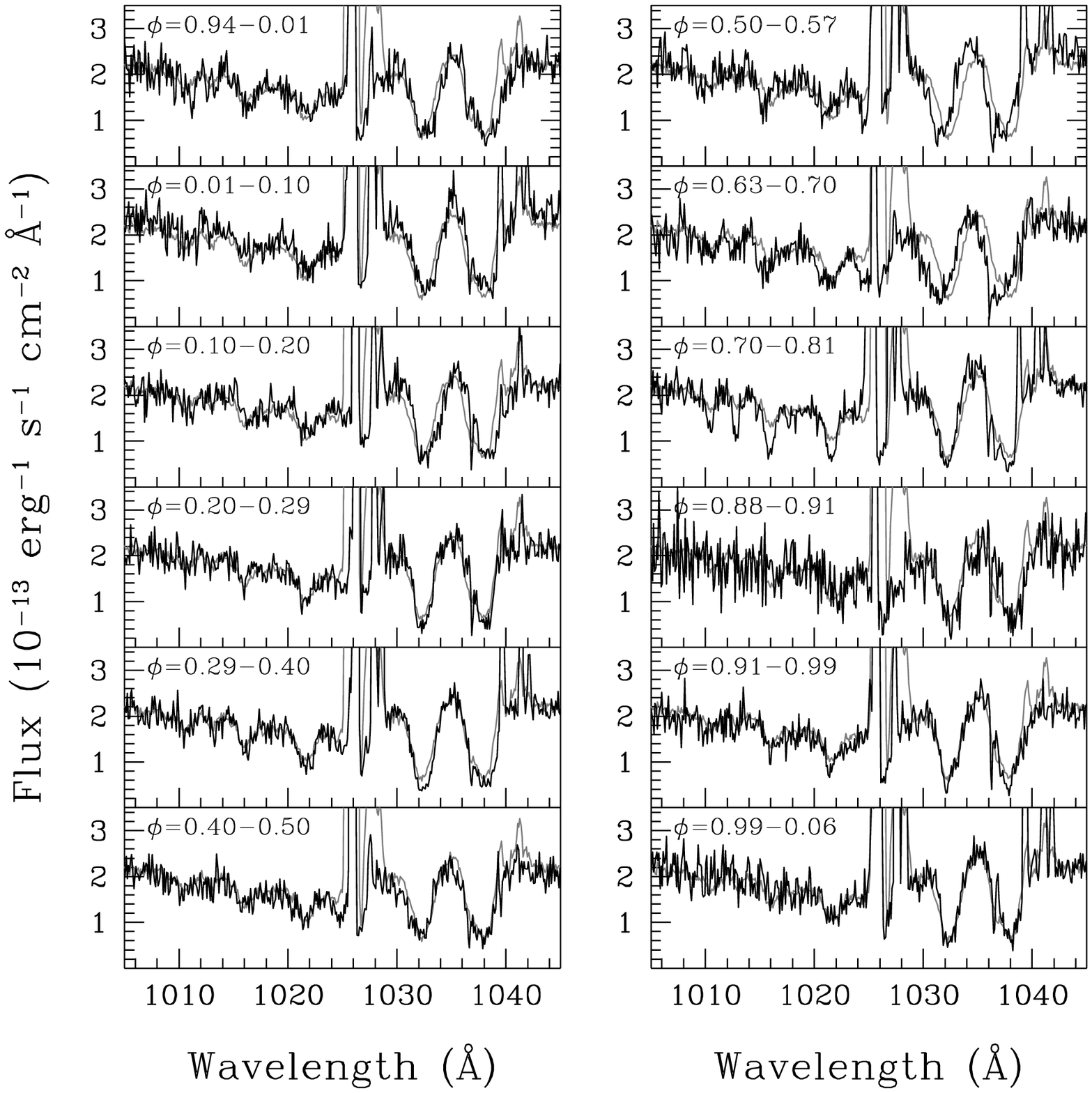}
\end{figure}

\begin{figure}
\plotone{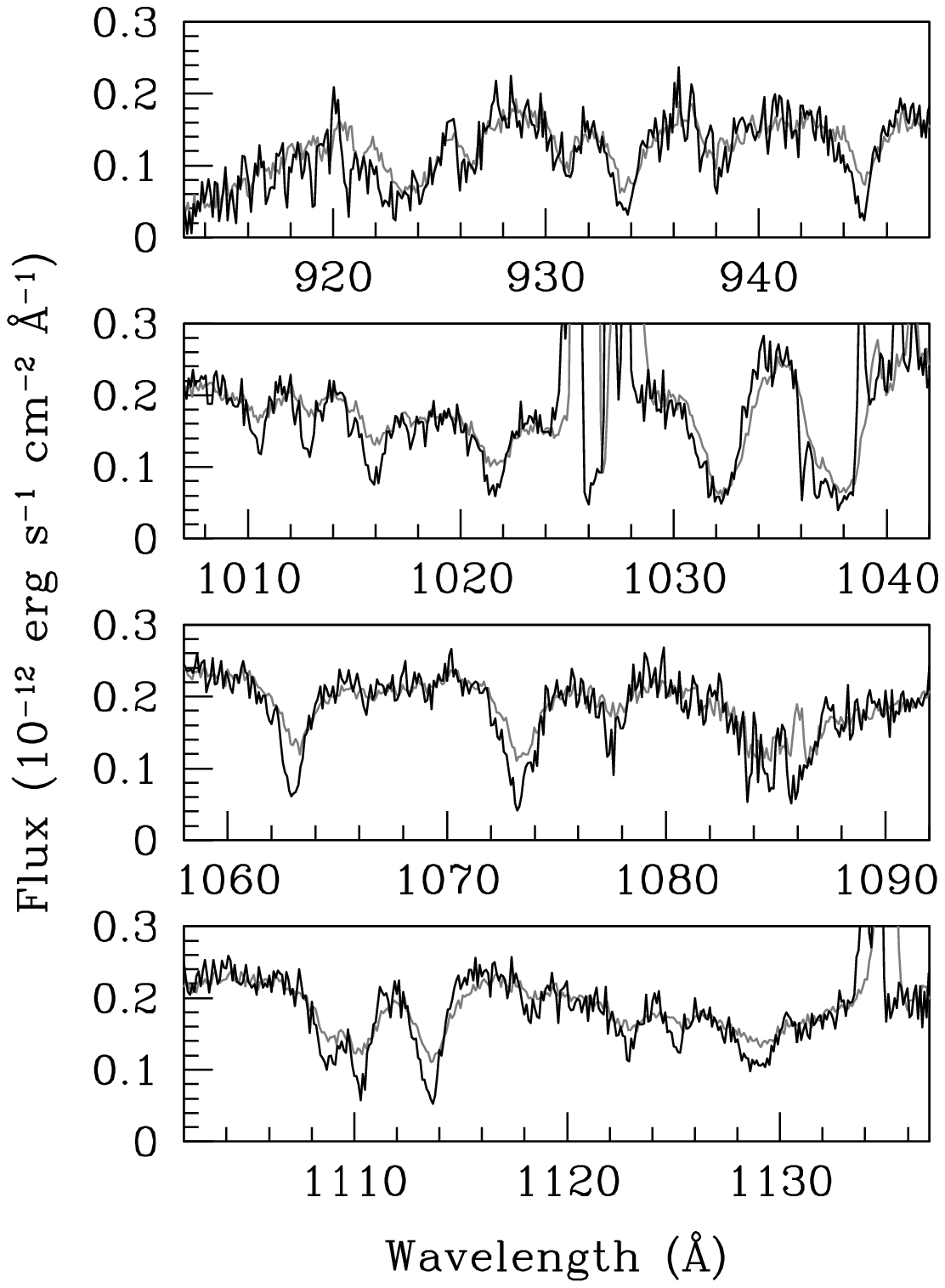}
\end{figure}

\begin{figure}
\plotone{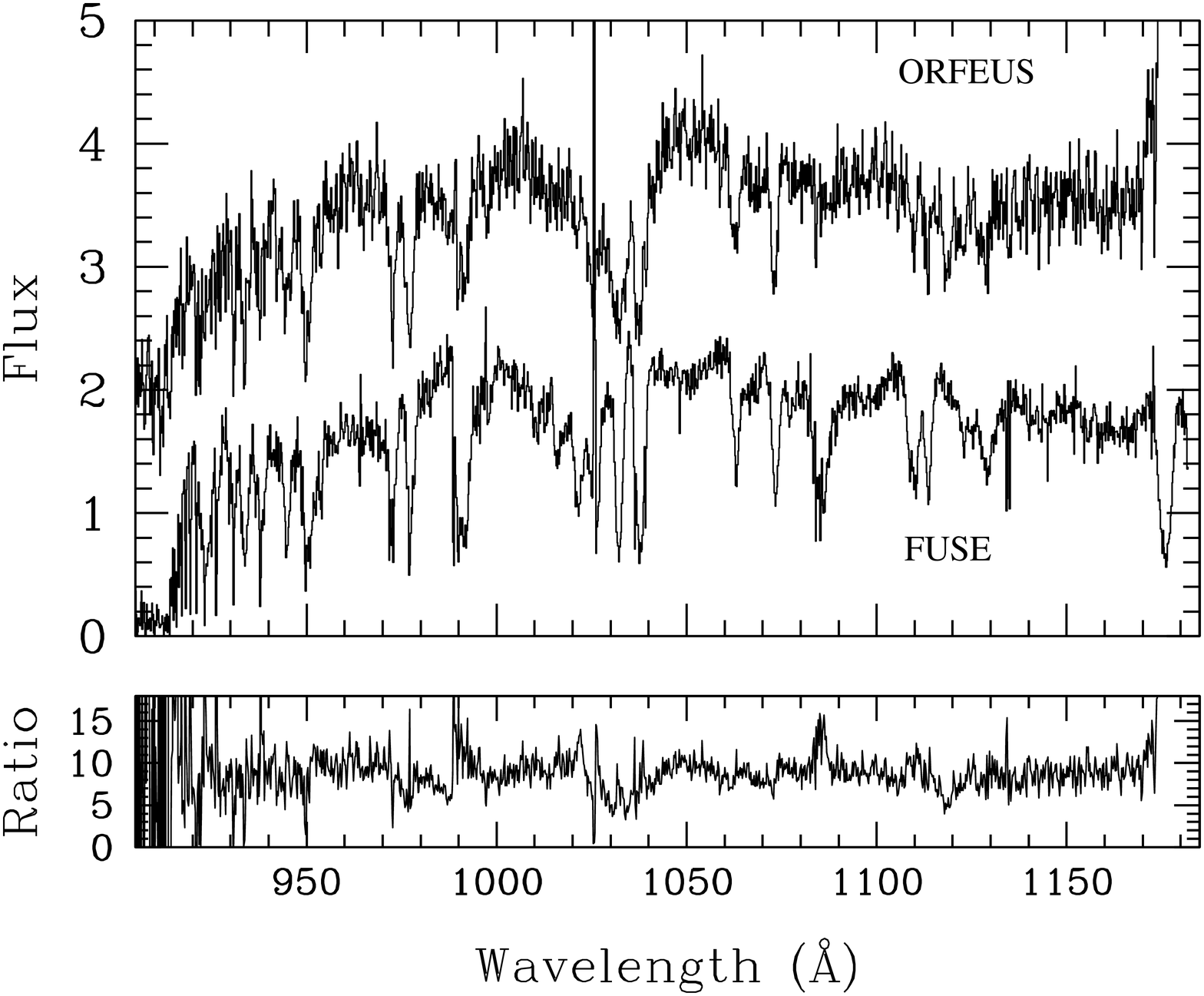}
\end{figure}

\begin{figure}
\plotone{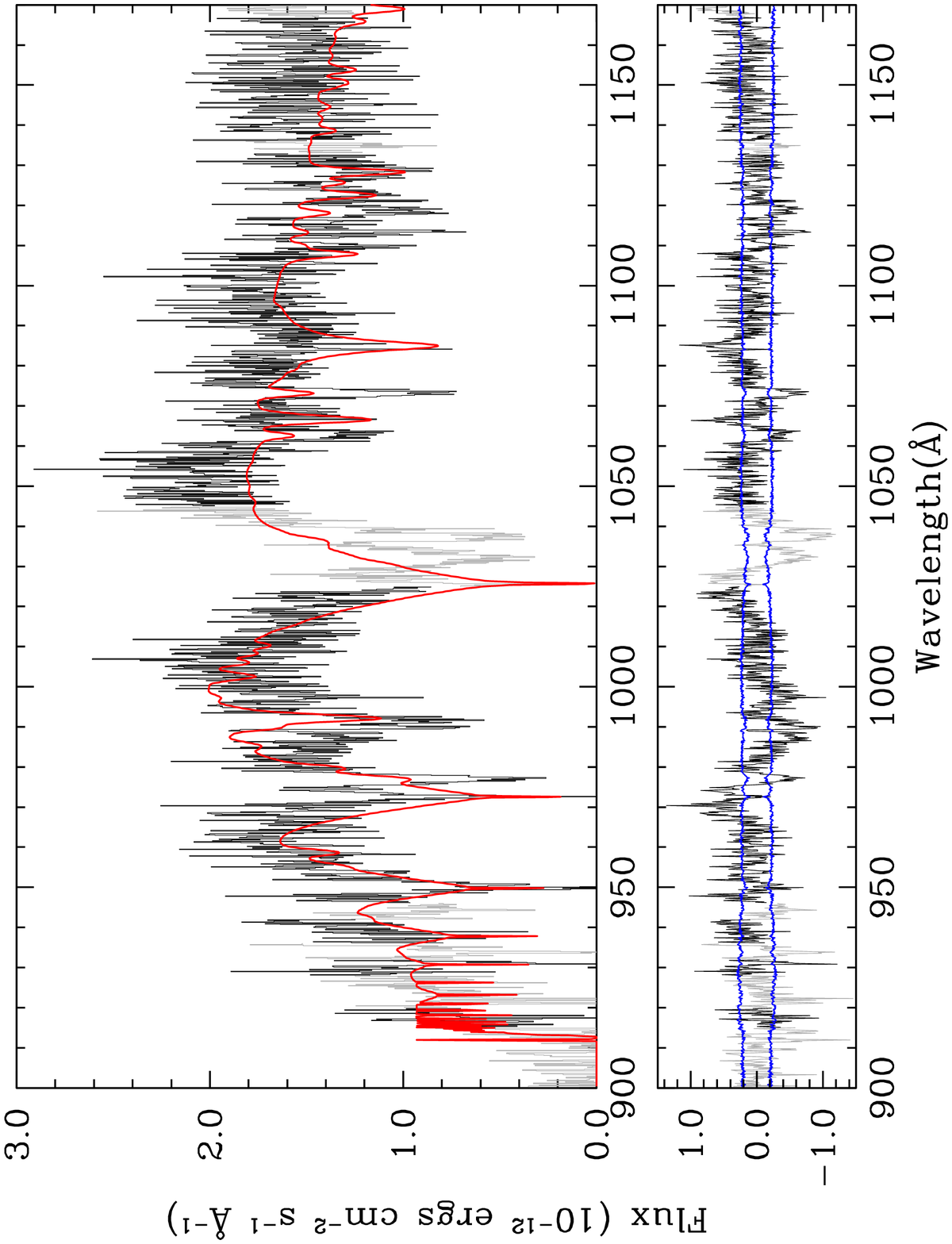}
\end{figure}

\begin{figure}
\plotone{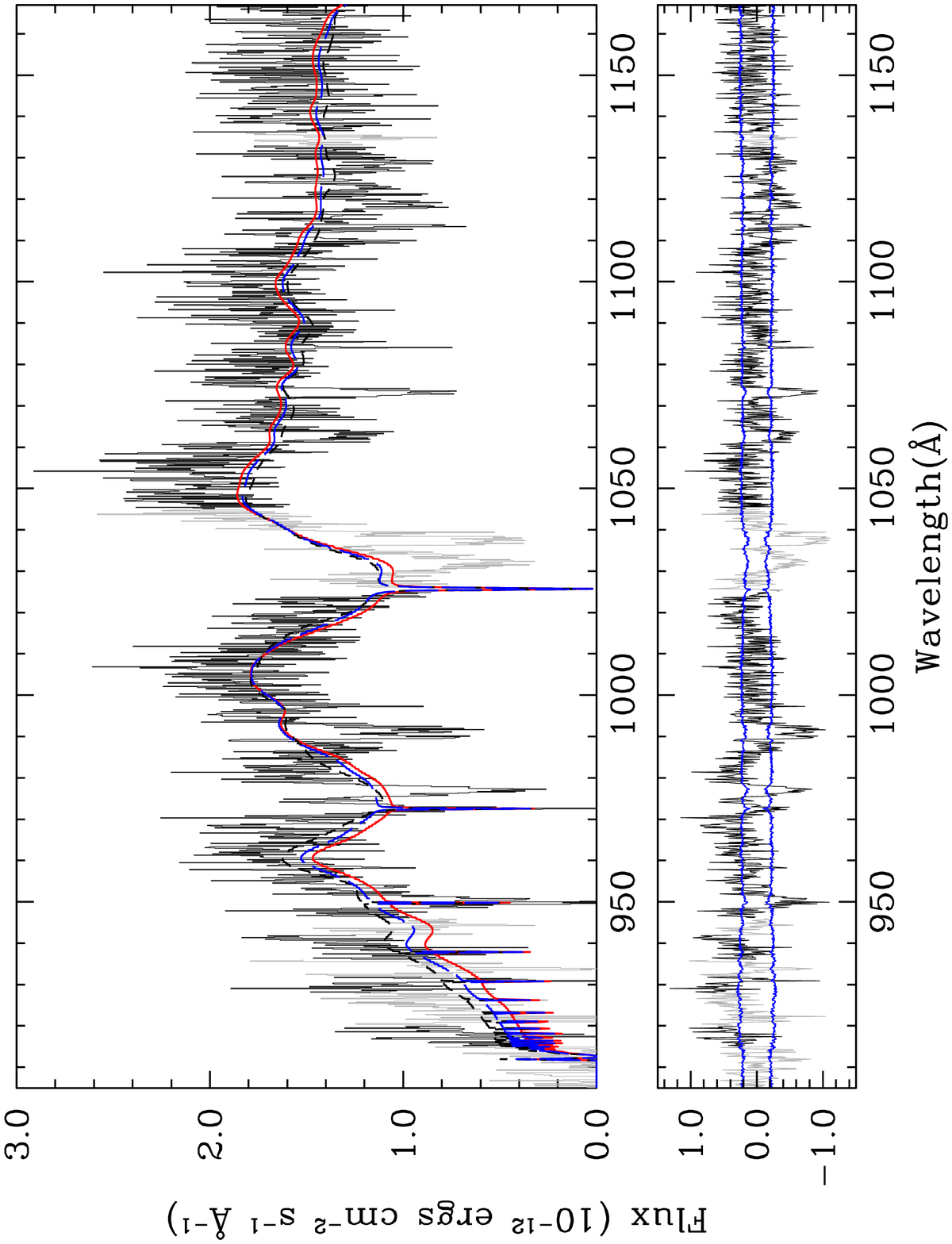}
\end{figure}

\end{document}